\documentclass[onecolumn,floats,floatfix,amssymb,nofootinbib,prd,superscriptaddress]{revtex4-1}

\usepackage{graphicx}
\usepackage{amsmath,amssymb}
\usepackage{amsfonts}
\usepackage{xspace} 
\usepackage[usenames]{color}
\usepackage{dcolumn}
\usepackage{bm}
\usepackage{mathrsfs}
\usepackage{multirow}
\usepackage[colorlinks=true]{hyperref}
\usepackage[all]{hypcap} 
\usepackage[utf8]{inputenc} 
\usepackage{lipsum}
\usepackage{etoolbox}
\usepackage{tikz}
\usepackage{url}
\usepackage{subcaption}
\usepackage{adjustbox}
\usepackage{caption}
\usepackage{slashed}
\usepackage{multirow}
\usepackage{caption}
\usepackage{array}
\usepackage{booktabs}
\usepackage{siunitx}
\usepackage{comment}
\usepackage{lineno}
\usepackage{todonotes}
\usepackage{gensymb}

\newtheorem{definition}{Definition}[section]
\newtheorem{note}{Note}[section]
\newtheorem{theorem}{Theorem}[section]
\newtheorem{corollary}{Corollary}[section]
\newtheorem{lemma}{Lemma}[section]
\newtheorem{proposition}{Proposition}[section]
\newtheorem{example}{Example}[section]

\newcommand{\bef}{\begin{definition}}
\newcommand{\eef}{\end{definition}}
\newcommand{\ben}{\begin{note}}
\newcommand{\een}{\end{note}}
\newcommand{\bep}{\begin{proposition}}
\newcommand{\eep}{\end{proposition}}
\newcommand{\bet}{\begin{theorem}}
\newcommand{\eet}{\end{theorem}}
\newcommand{\bee}{\begin{example}}
\newcommand{\eee}{\end{example}}
\newcommand{\bec}{\begin{corollary}}
\newcommand{\eec}{\end{corollary}}
\newcommand{\bel}{\begin{lemma}}
\newcommand{\eel}{\end{lemma}}

\def\be{\begin{equation}}
\def\ee{\end{equation}}
\def\bea{\begin{eqnarray}}
\def\eea{\end{eqnarray}}

\newcommand{\beq}{\begin{eqnarray}}
\newcommand{\eeq}{\end{eqnarray}}
\usepackage{calrsfs}
\DeclareMathAlphabet{\pazocal}{OMS}{zplm}{m}{n}

\DeclareMathAlphabet{\cmcal}{OMS}{cmsy}{m}{n}

\begin{document}

\title{Spontaneous particle creation by oscillating compact  stars}

\author{Adri\'an del R\'{\i}o}\email{adrdelri@math.uc3m.es}
\affiliation{Universidad Carlos III de Madrid, Departamento de Matem\'aticas.\\ Avenida de la Universidad 30 (edificio Sabatini), 28911 Legan\'es (Madrid), Spain. }
\author{Pau L\'opez-Oliver}
\email{pau.lopez.oliver@edu.unige.it}
\affiliation{Departamento de Fisica Teorica and IFIC, Centro Mixto Universidad de Valencia-CSIC. \\ Facultad de Fisica, Universidad de Valencia, Burjassot-46100, Valencia, Spain}
\affiliation{DIME, University of Genova, Via all'Opera Pia 15, 16145 Genova, Italy}
\affiliation{INFN Sezione di Genova, Via Dodecaneso 33, 16146 Genova, Italy}

\begin{abstract}

Quantum field theory predicts that dynamical curved spacetimes can spontaneously excite particle pairs from the quantum vacuum, a phenomenon  extensively studied in expanding universes and in scenarios involving gravitational collapse. In this article, we explore particle creation driven by radial oscillations of 3+1-dimensional spherically symmetric compact objects, such as neutron stars, using a massless, minimally coupled scalar field as a  reference model. We employ a toy model to describe the oscillatory dynamics and its coupling to the field modes, focusing on the resulting effects in the exterior spacetime of the star. The Bogoliubov coefficients relating the in and out vacua are computed non-perturbatively using high-precision numerical methods, without relying on weak-field, small-amplitude or small-velocity expansions. This allows us to determine the full particle spectrum and the total particle number in the strong-field and fully relativistic regime. Our  analysis confirms the existence of particle creation in this setting and, crucially, reveals a distinct resonance structure in the spectrum.

{\let\clearpage\relax  \let\sectionname\relax \tableofcontents}
\end{abstract}
\maketitle

\section{Introduction}
\label{introduction}

Strong gravitational fields can have a profound impact on quantum fields \cite{Birrell:1982ix, Fulling_1989, Wald:1995yp, Parker:2009uva, doi:10.1142/S0217751X13300238}, modifying the behavior of their vacuum fluctuations and enabling the physical excitation of particles by the spacetime itself. This phenomenon was first identified in the context of expanding universes through the seminal work of Parker \cite{parker2025creationparticlesexpandinguniverse, PhysRev.183.1057, PhysRevD.3.346}, and it now constitutes a cornerstone of modern cosmology. Namely, particle creation in time-dependent cosmological backgrounds provides the underlying mechanism responsible for the generation of primordial perturbations during the inflationary era \cite{Mukhanov:1981xt, HAWKING1982295, PhysRevLett.49.1110, STAROBINSKY1982175, PhysRevD.28.679, PhysRevD.89.084037}, which later seed the observed large-scale anisotropies of the Universe \cite{2020moco.book.....D}. An analogous mechanism plays a central role in black hole physics. As shown by Hawking \cite{Hawking:1974rv, cmp/1103899181}, the gravitational collapse of a star leading to black hole formation results in the emission of a late-time steady thermal flux of particles detectable by distant observers \cite{cmp/1103899393, PhysRevD.12.1519}, with a temperature determined by the properties of the final black hole. This remarkable connection between quantum field theory, gravitation, and thermodynamics has had far-reaching consequences, including the statistical interpretation of the black hole horizon area in terms of microscopic degrees of freedom, and has strongly influenced subsequent developments in quantum gravity research (see e.g. \cite{PhysRevLett.80.904, PhysRevLett.82.2828, doi.10.1142,   Wald:1999vt, Carlip_2000, PhysRevLett.100.211301, PhysRevD.83.104013} and references therein).

Particle creation effects are expected to become physically relevant only in regimes of strong gravity. For this reason, most previous studies have focused on early-universe cosmology (cosmic inflation) and on gravitational collapse, which provide the most natural and well-controlled settings. Nevertheless, there exist other physically relevant scenarios that have received comparatively little attention. One such setting is provided by astrophysical compact objects, such as neutron stars \cite{Haensel:2007yy}, or possibly more exotic ultracompact configurations \cite{Cardoso_2019}.  Compact stars undergoing relativistic oscillations play a central role in astrophysics, as their quasi-normal modes encode valuable information about the matter equation of state and the structure of spacetime in the strong-field regime \cite{Kokkotas:1999bd}. They therefore constitute natural laboratories for investigating the interplay between matter, gravity, and fundamental fields under extreme conditions, making them particularly relevant for the study of quantum effects in curved spacetimes.

Despite their physical relevance, this scenario has not yet been explored in connection with spontaneous particle creation, to the best of our knowledge. This is likely due to the significant technical challenges involved in the corresponding computations. In cosmological settings, for instance, the implications of particle creation can often be estimated using relatively simple numerical methods, or even derived analytically in closed form (see, e.g., \cite{Parker:2009uva} and references therein). In practice, the high degree of symmetry and conformal flatness of Friedmann-Lemaitre-Robertson-Walker spacetimes allow for a natural definition of the vacuum state \cite{PhysRevD.106.085003}. The problem then reduces to solving a single second-order ordinary differential equation (ODE) governing the time evolution of the field modes, with the gravitational dynamics encoded in a time-dependent effective frequency through the cosmological scale factor. Similarly, the thermal spectrum associated with Hawking radiation during gravitational collapse can be obtained in comparatively simple terms by focusing on the late-time Schwarzschild or Kerr geometry and neglecting the detailed structure of the collapsing star \cite{doi:10.1142/p378, delRio:2025ntl}.

In contrast, no comparable simplifications are expected to be available in the present problem. Modeling physically realistic neutron stars requires detailed knowledge of the matter equation of state and, in general, the numerical solution of the Einstein field equations for the dynamical spacetime geometry. Even for simplified polytropic equations of state commonly employed as toy models \cite{poisson2014gravity}, the gravitational field equations typically admit no closed-form solutions and must be treated numerically. Furthermore, the inclusion of stellar oscillations necessarily requires going beyond the stationary regime, leading to partial differential equations (PDEs) for the field modes and thereby substantially increasing the complexity of the problem.

On general grounds, the oscillations of neutron stars are expected to induce spontaneous particle creation through quantum effects. However, the detailed features of this process, including the spectrum of the created particles, remain largely unexplored and must be determined through explicit calculations. Recent work~\cite{Duarte-Baptista_2026} has investigated particle creation numerically within a toy model in which oscillations are simulated by effective time-dependent potentials in Minkowski spacetime. A corresponding analysis in genuinely curved spacetimes, however, is still lacking in the literature. The purpose of this article is to address this gap, at least partially. To simplify the analysis, we consider only  radial oscillations of a 3+1-dimensional spherically symmetric compact object and focus on their effects in the exterior region, where the symmetries uniquely fix the geometry to be Schwarzschild. In this approach, the detailed microphysics of the stellar interior is encoded through effective boundary conditions imposed on the field modes  at the stellar surface. For definiteness, we restrict attention to a massless, minimally coupled scalar field, and the interaction with the stellar matter is modeled through  Dirichlet boundary conditions. The oscillating matter is described by treating this stellar surface as a moving boundary, which induces a time-dependent modulation of the field modes and leads to the excitation of particle pairs after a finite duration of oscillatory motion, as expected from 1+1-dimensional models in flat space \cite{10.1098/rspa.1976.0045, 10.1098/rspa.1977.0130, Walker_1982, PhysRevD.31.767,  PhysRevD.36.2327, PhysRevD.88.025023, PhysRevD.111.065011}. Since our primary interest lies in the qualitative features of particle creation, we adopt a toy model for the stellar oscillation profile. A more complete treatment would require solving the full Einstein field equations numerically to obtain the dynamical oscillations of the spacetime metric.

Within this framework, our analysis is carried out in a genuinely strong-field setting: the 3+1-dimensional Schwarzschild geometry and the  field equations are treated exactly, without resorting to weak-field expansions or dimensional-reduction approximations. Furthermore, the stellar motion is treated in a fully relativistic manner, allowing for arbitrary oscillation amplitudes and without imposing slow-motion or small-acceleration approximations for the boundary trajectory. Finally, the Bogoliubov coefficients relating the in and out vacua are computed non-perturbatively using high-accuracy numerical methods. This framework enables us to effectively determine the particle spectrum across the full range of dynamically relevant modes, including off-resonant mode-mixing contributions, and to compute the total particle number.

It is worth noting that, although oscillations are used as a representative example due to their relevance in astrophysics, the procedure developed in this paper applies to any radial process that is asymptotically static in both the far past and the far future. Similarly, the use of Dirichlet boundary conditions at the stellar surface is not essential: our framework can be straightforwardly extended to a wider class of boundary conditions.

The article is organized as follows. In Sec.~\ref{framework} we introduce the spacetime background and review the elements of quantum field theory in curved spacetimes relevant to particle creation. We also present the toy model for the stellar oscillations and discuss our method for solving the field equations with time-dependent boundary conditions. Section~\ref{numerical_methods} describes the numerical methods employed in the analysis. The resulting particle spectrum and total particle number are presented in Sec.~\ref{results}. Finally, Sec.~\ref{conclusions} summarizes our findings and discusses possible extensions of this work.

Throughout this paper we employ natural geometrized units with $c = G = 1$ while keeping $\hbar\neq 1$ to emphasize quantum effects. We  adopt the mostly-plus metric signature $(-+++)$ for the spacetime metric $g_{a b}$. The covariant derivative associated with the Levi-Civita connection is denoted by $\nabla_a$, and $R^a_{\,\, \,bcd }$ denotes the corresponding Riemann  curvature tensor.

\section{Theoretical framework}
\label{framework}

In this section we specify the spacetime background and the quantum field considered in this work. We also review basic aspects of field quantization in curved spacetimes, with particular emphasis on the well-known ambiguity in the definition of the vacuum state and its physical implications for spontaneous particle creation. This section additionally serves to fix notation and conventions.

\subsection{Spacetime background}

Modeling the spacetime interior of physically realistic compact stars is technically challenging, as it requires specifying reasonable equations of state for the  matter content \cite{Haensel:2007yy}. While polytropic models often yield reasonable qualitative results, they still necessitate the use of numerical methods to solve the Einstein's field equations for the spacetime metric \cite{poisson2014gravity}. Furthermore, stellar oscillations are typically treated perturbatively \cite{PhysRevLett.12.114,1964ApJ...140..417C,Kokkotas2001}, and consistently incorporating their backreaction into Einstein's equations to obtain the full time-dependent metric components introduces an additional layer of computational complexity. Since the primary goal of this work is to investigate the qualitative features of particle creation in dynamical stars, we restrict attention to the exterior region of the star and model the interaction between the quantum field and the stellar matter through effective boundary conditions imposed at the stellar surface, to be specified later.

Accordingly, we  consider the  exterior spacetime $(\mathbb M,g_{ab})$ of a spherically symmetric star of mass $M$.  The four-dimensional manifold $\mathbb M$ is diffeomorphic to $\mathbb R\times (R_0,\infty)\times \mathbb S^2$, where  $R_0>2M$ denotes the initial radius of the star, while  the metric $g_{ab}$ satisfies the vacuum Einstein field equations, $R_{ab}[g]=0$. By Birkhoff's theorem \cite{Wald:1984rg}, the spacetime geometry is uniquely given by the Schwarzschild line element
\bea
ds^2=g_{ab}dx^a dx^b=-f(r) dt^2+ f(r)dr^2 + r^2 \left( d\theta^2+ \sin^2\theta d\phi^2\right)\, , \label{metric}
\eea
where $f(r):=1-\frac{2M}{r}$. The  coordinates take  values  $(t,\theta,\phi)\in \mathbb R\times (0,\pi)\times (0,2\pi)$, while the radial coordinate is restricted to $r\in(R_0(t),\infty)$, where the function $R_0(t)$ describes the time-dependent radius of the stellar surface.

The spacetime $(\mathbb{M}, g_{ab})$ possesses a timelike boundary $\partial\mathbb{M}$ defined by the stellar surface \cite{Ak_Hau_2020}. Notice that all the dynamics is encoded in this boundary. We consider a scenario in which the star is initially at rest, undergoes a finite period of oscillatory motion starting at some time $t_{\rm on}$, and subsequently relaxes back to a static configuration at a later time $t_{\rm off} > t_{\rm on}$. To preserve spherical symmetry at all times, we restrict attention to periodic {\it radial} oscillations, such that the stellar radius remains independent of the angular coordinates $(\theta,\phi)$. In this setting, points on the stellar surface follow timelike worldlines of the form $\gamma_{\theta,\phi}(t)=(t,R_0(t),\theta,\phi)$ for each $(\theta,\phi) \in \mathbb{S}^2$. The function $R_0(t)$ interpolates between radii $R_0$ and $R_1$ during the oscillatory phase. For simplicity, and without loss of generality for our purposes, we set $R_1 = R_0$.

\subsection{Field quantization}

We now  consider  a free massless, minimally coupled  scalar field  on this spacetime  $(\mathbb M,g_{ab})$. The  dynamics of the classical field $\phi$ is governed by the homogeneous Klein-Gordon equation,
\bea
g^{ab}\nabla_a \nabla_b \phi(x)=0\, . \label{kg}
\eea
Since the spacetime background is spherically symmetric, the most general solution can be expanded in normal modes as
\bea
\phi(t,r,\theta,\phi)=\sum_{\ell=0}^{\infty}\sum_{m=-\ell}^{\ell} R_{\ell}(t,r)Y_{\ell m}(\theta,\phi)\, ,
\eea
where $Y_{\ell m}$ denote the standard spherical harmonics, and $R_{\ell}(t,r)$ satisfies the second-order linear PDE\footnote{With the field redefinition $R_\ell\to \frac{1}{r}R_\ell$, this equation reduces to a time-dependent Regge-Wheeler equation \cite{Chandrasekhar:579245}: $[-\partial_t^2+\partial^2_{r^*} -V_\ell(r)]R_{\ell}(r)=0$, where $r_*=r+2M \log \left(\frac{r}{2M}-1\right)$ denotes the tortoise coordinate and $V_\ell(r)=f(r)\frac{\ell(\ell+1)\tau+2M}{\tau^3}$ an effective potential barrier centered around $r\sim 3M$.}
\bea
 \left[ - \frac{\partial^{2}}{\partial t^{2}} +  f(r)^{2} \dfrac{d^{2}}{d r^{2}} + \frac{f(r)}{r^2}\partial_r(r^2 f(r))  \dfrac{d}{d r} - \dfrac{\ell (\ell + 1) f(r)}{r^{2}} \right] R_{\ell}(t, r) = 0\, .  \label{ode}
\eea
The covariant phase space $\Gamma$ of the classical theory \cite{10.1063/1.528801, PhysRevD.103.025011, PhysRevD.105.L101701} is defined as the real vector space spanned by all smooth solutions of Eq.~(\ref{ode}), subject to appropriate boundary conditions. These conditions are dictated by the physical setting, namely the interaction of the field modes with the stellar surface and suitable fall-off behavior at large radial distances. 

To effectively model the coupling between the quantum field and the stellar matter, we impose Dirichlet boundary conditions at the stellar surface, $R_{\ell}(t, R_0(t))=0$. Physically, the star thus acts as an effective mirror for field modes propagating in the exterior vacuum.   While simple, this toy model suffices to capture the  qualitative features relevant for particle creation, and  is a commonly adopted idealization in the literature on compact objects \cite{Cardoso_2019}. The Dirichlet condition can be viewed as an effective potential barrier at the stellar surface, arising from the abrupt change in matter density at the stellar interface \cite{Volkel_2019, PhysRevD.101.104035}. 
For $\omega M\lesssim 1$, scattering analyses show that massless waves become largely insensitive to the stellar interior and the response is dominated by the exterior curvature potential \cite{PhysRevD.95.124055, PhysRevD.100.024007}. In this regime, modeling the stellar surface as effectively reflecting provides a reasonable phenomenological approximation. Since typical stellar oscillation frequencies lie in the range $O(0.01 M^{-1}) - O(0.1 M^{-1})$, this constitutes a well-motivated phenomenological boundary condition in our problem. On the other hand, since the spacetime metric is asymptotically flat, i.e.\ $g_{ab} \to \eta_{ab}$ as $r \to \infty$, we require the field modes to asymptotically approach their Minkowski counterparts. In particular, for large $r$ we impose $R_{\ell}(t, r)\sim\int_{-\infty}^{\infty} d\omega j_\ell(\omega r)e^{-i\omega t}\sim O(r^{-1})$ as $r\to\infty$, where $j_\ell$ denotes the spherical Bessel function of order $\ell$.

This vector space $\Gamma$ can be endowed with a canonical symplectic structure given by 
\bea
\Omega(\phi_1,\phi_2)=\int_{\Sigma_t} \,d^3\vec x\sqrt{h}  \, (\phi_{1} \,  t^a\nabla_a  \phi_2- \phi_{2} \, t^a\nabla_a \phi_1)\, , \label{symp}
\eea
where $t^a = -\left(-g^{tt}\right)^{-1/2} g^{ab} \nabla_b t$ is the future-directed unit timelike normal to the spacelike hypersurfaces $\Sigma_{t_0} = \{ p \in \mathbb{M} \mid t(p) = t_0 \}$, and $h$ denotes the determinant of the induced metric on $\Sigma_{t_0}$. Owing to the field equation (\ref{kg}) and the imposed boundary conditions, this symplectic form is conserved in time $t$ for any $\phi_1,\phi_2 \in \Gamma$ (see Appendix \ref{app:time-independence-symplectic}).

Quantization can then be performed using standard algebraic techniques \cite{Ashtekar1975zn, doi:10.1142/S0217751X13300238}. To each classical solution $\phi \in \Gamma$ we associate an abstract element $\mathcal O(\phi)$, called the field operator, and construct the free $*$-algebra $\mathcal{A}$ generated by these symbols subject to the following relations: (i) linearity in $\phi$, (ii) hermiticity, and (iii) the covariant commutation relations $[\mathcal O(\phi_1),\mathcal O(\phi_2)]=i\,\hbar\, \Omega(\phi_1,\phi_2)\, \mathbb I$. To obtain a Hilbert-space representation of this algebra, one can introduce a complex structure $J$ on $\Gamma$, i.e. a linear map $J : \Gamma \to \Gamma$, satisfying $J^2 = -\mathbb{I}$, which allows us to decompose $\Gamma$ as a direct sum of two subspaces, $\Gamma\simeq \Gamma^+\oplus \Gamma^-$. More precisely, we say that $\phi^+=\frac{1-i J}{2}\phi\in \Gamma^+$ is the positive-frequency part of $\phi$, while  $\phi^-=\frac{1+i J}{2}\phi\in \Gamma^-$ is the negative-frequency part, both  satisfying $J\phi^\pm=\pm i \phi^\pm$ so that $J\phi=i\phi^+-i\phi^-$. Then, if $J$ is compatible with the symplectic product (\ref{symp}), in the sense that $\Omega(J\phi_1,J\phi_2)=\Omega(\phi_1,\phi_2)$,  this $J$ naturally leads to a hermitian inner product, hereafter referred to as the Klein-Gordon inner product, defined by
\bea
\langle \phi_1, \phi_2\rangle=\frac{1}{2\hbar}\left[ \Omega(\phi_1, J\phi_2)+i \Omega(\phi_1, \phi_2)\right]\, , \label{innerproduct}
\eea
which converts the completion of  $(\Gamma, J, \langle \cdot,\cdot \rangle)$ into the 1-particle Hilbert space $\mathcal H$ of the theory.  The full space of quantum states is then the symmetric Fock space based on this Hilbert space, $\mathcal{F} = \bigoplus_{n=0}^{\infty} S\,\mathcal{H}^{\otimes n}$.

In a generic curved spacetime there is no preferred choice of complex structure $J$, reflecting the physical ambiguity in the definition of the vacuum state in quantum field theory. However, when the spacetime is stationary, admitting  a timelike Killing vector field $t^a$, a natural choice of $J$ can be constructed. This applies to our setup precisely at early and late times. To see this, let us Fourier-decompose any $\phi \in \Gamma$ with respect to the proper time of $t^a$,
\bea
\phi(t,r,\theta,\phi)=\int_{-\infty}^{\infty}d\omega\, \phi_\omega(r,\theta,\phi)\, e^{-i\omega t}=\int_{-\infty}^{\infty}d\omega\, \sum_{\ell=0}^{\infty}\sum_{m=-\ell}^{\ell}\, e^{-i\omega t}R_{\omega\ell }(r)Y_{\ell m}(\theta,\phi)\, ,
\eea
where the reality condition gives $\phi_{-\omega}=\overline{\phi_\omega}$. Then,  one can define $J=-\theta^{-1/2}\mathcal L_t$, where $\theta=-\mathcal L_t \mathcal L_t$ with $\mathcal L$ denoting the Lie derivative  \cite{Ashtekar1975zn}. This construction yields $J\phi=i\int_{-\infty}^{\infty}d\omega\, {\rm sign}\, \omega\, \phi_\omega(r,\theta,\phi)\, e^{-i\omega t}=i\phi^+-i\phi^-$, where $\phi^+=\int_{0}^{\infty}d\omega\,  \phi_\omega(r,\theta,\phi)\, e^{-i\omega t}$ is  the frequency part of $\phi$, while $\phi^-=\int_{0}^{\infty}d\omega\,  \overline{\phi_{\omega}(r,\theta,\phi)}\, e^{i\omega t}=\overline{\phi^+}$ gives the negative-frequency component. In particular, $\phi=\phi^++\phi^-$. Now, if  the modes $\{\phi_\omega\}_{\omega\in \mathbb R}$ are orthogonal with respect to the $L^2$ inner product on $\Sigma_t$, $\int d^3x \sqrt{h} \phi_\omega \phi_{\omega'}\propto \delta(\omega-\omega')$, then it can be checked that this construction is compatible with the symplectic structure given in equation (\ref{symp}), namely $\Omega(J\phi_1,J\phi_2)=\Omega(\phi_1^+,\phi_2^-)+\Omega(\phi_1^-,\phi_2^+)=\Omega(\phi_1,\phi_2)$. Furthermore, the hermitian inner product (\ref{innerproduct}) reduces to the simple expression
\bea
\langle \phi_1, \phi_2\rangle=\frac{i}{\hbar}\Omega(\phi_1^-,\phi_2^+)=\frac{i}{\hbar}\int_{\Sigma_t} \,d^3\vec x\sqrt{h}  \, (\overline{\phi^+_{1}} \,  t^a\nabla_a  \phi_2^+- \phi_{2}^+ \, t^a\nabla_a \overline{\phi_1^+})\, .\label{inner2}
\eea
which in turn endows the complex space $\Gamma^+$ with the inner product $\langle \phi^+_1, \phi^+_2\rangle_+:=\frac{i}{\hbar}\Omega(\overline{\phi_1^+},\phi_2^+)$.

In the quantum theory the field operator $\mathcal O(\phi)$ can be represented in terms of an  operator-valued distribution $\hat\phi(x)$ smeared with a test function $f\in C^{\infty}_0(\mathbb M)$, via $\mathcal O(\phi)=\int d^4x \sqrt{-g}\, \hat\phi(x)f(x)$. In practical computations it is more convenient to work with $\hat\phi(x)$ directly. This    can be expanded  in creation  and annihilation operators, $a_{\omega\ell m}$ and $a^{\dagger}_{\omega\ell m}$ respectively,   in the usual form
\bea
\hat \phi(x) =  \int_0^{\infty}d\omega \sum_{\ell=0}^{\infty}\sum_{m=-\ell}^{+\ell} \left[a_{\omega\ell m}\, \phi^+_{\omega\ell m}(x) +a^{\dagger}_{\omega\ell m}   \overline{\phi^+_{\omega\ell m}}(x) \right]\, ,\label{field}
\eea
where the field modes $\phi^+_{\omega\ell m}(x)$ are the solutions of (\ref{kg}) of positive-frequency with respect to the chosen $J$, labeled by  the quantum numbers $\{\omega,\ell,m\}$. This field operator acts on the Fock space of quantum states in the usual way. The vacuum or ground state of the theory is defined by  $a_{\omega\ell m}|0\rangle=0$ for any $\omega,\ell,m$, and the rest of states of the Fock space can be obtained by acting successively with the creation operator $a^{\dagger}_{\omega\ell m}$ on $|0\rangle$. The creation  and annihilation operators  satisfy $[a_{\omega\ell m},a^{\dagger}_{\omega'\ell' m'}]=\delta(\omega-\omega')\delta_{\ell,\ell'}\delta_{m,m'}$, $[a^{\dagger}_{\omega\ell m},a^{\dagger}_{\omega'\ell' m'}]=0$, $[a_{\omega\ell m},a^{}_{\omega'\ell' m'}]=0$, which ensure the equal-time canonical commutation relations for the field, $[\hat\phi(t,\vec x), \hat\Pi(t,\vec y)]=i\hbar\,  \delta^{(3)}(\vec x- \vec y)$, on any $\Sigma_t$.

\subsection{In and out vacua and particle creation}

Unlike ordinary quantum mechanics, in quantum field theory there is an inherent ambiguity in the vacuum  state of the theory, ultimately rooted in the presence of infinitely many physical degrees of freedom. As described in the previous subsection, constructing a Hilbert-space  representation of the canonical commutation relations requires introducing  a complex structure $J$ on the  covariant phase space $\Gamma$, but in a generic  spacetime there exist infinitely many admissible choices. In practice, fixing this freedom is equivalent to specifying a unique solution to the PDE problem (\ref{ode}). Namely, different solutions of (\ref{ode}) for the positive-frequency field modes $\phi^+_{\omega\ell m}(x)$ in the expansion (\ref{field})  can lead to different (potentially physically inequivalent) notions of $|0\rangle$ through the associated annihilation operators. This ambiguity is fully  fixed only when the field modes $\phi_{\omega\ell m}(x)$ are univocally specified as solutions of the field equation (\ref{kg}) by e.g. prescribing appropriate data on a Cauchy hypersurface  $\Sigma_t$.

In our  setting, the spacetime background is stationary  at both early and late times. Consequently, one can single out preferred complex structures $J_{\rm in}$ and $J_{\rm out}$ associated with the timelike Killing symmetry in each asymptotic regime. This yields two natural representations  of the field operator (\ref{field}) in terms of creation and annihilation operators, each one associated to early (``in'') and late (``out'') times,
\bea
\hat \phi(x) =  \int_0^{\infty}d\omega \sum_{\ell=0}^{\infty}\sum_{m=-\ell}^{+\ell} \left[a^{\rm in}_{\omega\ell m}\, \phi^{\rm in}_{\omega\ell m}(x) +a^{{\rm in}, \dagger}_{\omega\ell m}   \overline{\phi^{\rm in}_{\omega\ell m}}(x) \right]\, ,\label{in}\\
\hat \phi(x) =  \int_0^{\infty}d\omega \sum_{\ell=0}^{\infty}\sum_{m=-\ell}^{+\ell} \left[a^{\rm out}_{\omega\ell m}\, \phi^{\rm out}_{\omega\ell m}(x) +a^{{\rm out}, \dagger}_{\omega\ell m}   \overline{\phi^{\rm out}_{\omega\ell m}}(x) \right]\, .\label{out}
\eea
In each case, the corresponding basis of positive-frequency field modes are uniquely determined, respectively, by the initial and ``final'' data:
\bea
 \phi^{\rm in}_{\omega\ell m}(x)= \sqrt{\hbar}\, A_{\omega\ell} R^{\rm in}_{\omega\ell}(t,r)Y_{\ell m}(\theta,\phi)\sim \sqrt{\hbar}\, A_{\omega\ell}   e^{-i\omega t} R_{\omega\ell}(r)Y_{\ell m}(\theta,\phi)\, , &\quad& {\rm for}\quad t< t_{\rm on}\, ,\label{inearly}\\
 \phi^{\rm out}_{\omega\ell m}(x) =\sqrt{\hbar}\, A_{\omega\ell}  R^{\rm out}_{\omega\ell}(t,r)Y_{\ell m}(\theta,\phi)\sim \sqrt{\hbar}\, A_{\omega\ell}   e^{-i\omega t} R_{\omega\ell}(r)Y_{\ell m}(\theta,\phi)\, , &\quad& {\rm for}\quad  t> t_{\rm off} \,  , \label{outlate}
\eea
where   the radial modes $R_{\omega\ell}(r)$ satisfy the static version of (\ref{ode}),
 \bea
 \left[   f(r)^{2} \dfrac{d^{2}}{d r^{2}} + \frac{f(r)}{r^2}\partial_r(r^2 f(r))  \dfrac{d}{d r} +\omega^2- \dfrac{\ell (\ell + 1) f(r)}{r^{2}} \right] R_{\omega\ell}(t, r) = 0\, , \label{ode2}
\eea
and $A_{\omega\ell}$ is a normalization factor chosen such that $\langle  \phi^{\rm in/out}_{\omega\ell m}, \phi^{\rm in/out}_{\omega'\ell' m'}\rangle_+=\delta(\omega-\omega')\delta_{\ell \ell'}\delta_{m m'}$. 
These two representations of the quantum field (\ref{in})-(\ref{out}) give rise to two different ``in'' and ``out'' vacuum states in the theory: $a^{\rm in}_{\omega\ell m}|{\rm in}\rangle=0$, $a^{\rm out}_{\omega\ell m}|{\rm out}\rangle=0$, respectively. For convenience, we choose the radial functions $R_{\ell}(r)$ to be dimensionless, so that the normalization constant $A_{\omega\ell}$ has dimensions of $\text{length}^{-1/2}$.

If $\mathcal F_{\rm out}\neq \mathcal F_{\rm in}$, the in vacuum state is seen as a state of particles in the out Fock space.  This is the phenomenon of spontaneous particle creation.  
To make this explicit, one must relate the two vacua. Since both $\{ \phi^{\rm in}_{\omega\ell m},\overline{\phi^{\rm in}_{\omega\ell m}}\}_{\omega\ell m}$ and $\{ \phi^{\rm out}_{\omega\ell m}, \overline{\phi^{\rm out}_{\omega\ell m}}\}_{\omega\ell m}$  form a complete orthonormal basis of $\Gamma$, it is possible to express  vectors of each basis as a linear combination of the vectors of the other basis:
\bea
\phi^{\rm in}_{\vec k}(x) &=&\sum_{\vec k'} \left[\alpha_{\vec k\vec k'}\phi^{\rm out}_{\vec k'}(x)+ \beta_{\vec k\vec k'}\overline{\phi^{\rm out}_{\vec k'}}(x)\right]\, ,\label{bfield}\\
\overline{\phi^{\rm in}_{\vec k}}(x) &=&\sum_{\vec k'} \left[\overline{\alpha_{\vec k\vec k'}} \overline{\phi^{\rm out}_{\vec k'}}(x)+ \overline{\beta_{\vec k\vec k'}} \phi^{\rm out}_{\vec k'}(x)\right]\, ,
\eea
where $\vec k$, $\vec k'$ are shorthands for $(\omega,\ell,m)$, $(\omega',\ell',m')$, respectively,  $\sum_{\vec k}\equiv  \int_0^{\infty}d\omega \sum_{\ell=0}^{\infty}\sum_{m=-\ell}^{+\ell}$, and $\alpha_{\vec k\vec k'}$, $\beta_{\vec k\vec k'}$ are some complex numbers known as Bogoliubov coefficients.
 Substituting these two expressions in (\ref{in}), and comparing with (\ref{out}), we can obtain  a similar linear relation between the creation and annihilation operators of the two representations, known as Bogoliubov transformation \cite{parker2025creationparticlesexpandinguniverse, birrell_davies_1982}:
\bea
a^{\rm out }_{\vec k}&=&\sum_{\vec k'}^{} \left(\overline{\alpha_{\vec k' \vec k}} a^{\rm in}_{\vec k'}-\overline{\beta_{\vec k' \vec k}} a_{\vec k'}^{\rm in, \dagger}\right)\, ,\label{Bogoannihilation}\\
a^{\rm out,\dagger }_{\vec k}&=&\sum_{\vec k'}^{} \left(\alpha_{\vec k' \vec k} a^{\rm in,\dagger}_{\vec k'}-\beta_{\vec k' \vec k} a_{\vec k'}^{\rm in}\right)\, .
\eea
It follows from (\ref{Bogoannihilation}) that $|\rm in\rangle \neq |\rm out\rangle$ if and only if $\beta_{\vec k\vec k'}\neq 0$. This is the key quantity that quantifies the effect of particle creation. Specifically, the number density of particles with quantum numbers $\vec k$, created out of the ``in'' vacuum by the oscillating star during the time $t_{\rm off}-t_{\rm on}$,  is given  by 
\bea
n^{\rm out}_{\vec k}=\langle{\rm in}|a_{\vec k}^{\rm out, \dagger} a^{\rm out}_{\vec k}| {\rm in}\rangle= \sum_{\vec k'}^{} \left|\beta_{\vec k'\vec k}\right|^{2}\, .
\eea
Then, the expected total number of particles created at late times in the state ``in'' is
\bea
\langle {\rm in}|N_{\rm out}|{\rm in}\rangle= \sum_{\vec k}^{}\sum_{\vec k'}^{} \left|\beta_{\vec k \vec k'}\right|^{2}\, . \label{particlecreation}
\eea
Since $\langle {\rm in}|N_{\rm out}|{\rm in}\rangle\neq 0$ if and only if $\beta_{\vec k, \vec k'}\neq 0$ for some $\{\vec k,\vec k'\}$, we conclude that $|\rm in\rangle \neq |\rm out\rangle$ if and only if there exists particle creation. Because  time evolution is driven via a unitary operator, the two Fock representations are unitarily equivalent precisely when the above expectation value is finite, $\langle {\rm in}|N_{\rm out}|{\rm in}\rangle < \infty$; otherwise they are unitarily inequivalent.

Our ultimate goal is to determine the frequency spectrum $\left|\beta_{\vec k \vec k'}\right|$ and to evaluate (\ref{particlecreation}). The Bogoliubov coefficient $\beta_{\vec k\vec k'}$ can be easily extracted from (\ref{bfield}) using the antisymmetry of the symplectic structure. From the orthonormal properties of the in and out modes, $\Omega (\phi^{\rm in/out}_{\omega\ell m},\overline{ \phi^{\rm in/out}_{\omega'\ell' m'}})=i\hbar\,\delta(\omega-\omega')\delta_{\ell \ell'}\delta_{m m'}$, one finds
\bea
\beta_{\vec k \vec k'}=\frac{\Omega(\phi^{\rm out}_{\vec k'},  \phi^{\rm in}_{\vec k})}{i\hbar} &=&\frac{i}{\hbar} \int_{R_0(t)}^{\infty} dr r^2\left(1-\frac{2M}{r}\right)^{-1} d\cos\theta\, d\phi\, \left[\phi^{\rm in}_{\vec k}\partial_t \phi^{\rm out}_{\vec k'} - \phi^{\rm out}_{\vec k'}\partial_t \phi^{\rm in}_{\vec k} \right] \, .
 \label{betacoefficient}
\eea
The $\alpha$ coefficients can be obtained in a completely analogous manner 
\bea
\alpha_{\vec k \vec k'}=\frac{\Omega(\overline{\phi^{\rm out}_{\vec k'}},  \phi^{\rm in}_{\vec k})}{-i\hbar} &=&-\frac{i}{\hbar} \int_{R_0(t)}^{\infty} dr r^2\left(1-\frac{2M}{r}\right)^{-1} d\cos\theta\, d\phi\, \left[\phi^{\rm in}_{\vec k}\partial_t \overline{\phi^{\rm out}_{\vec k'}} - \overline{\phi^{\rm out}_{\vec k'}}\partial_t \phi^{\rm in}_{\vec k} \right] \, .
 \label{betacoefficient}
\eea
We note that both the $\alpha$ and $\beta$ coefficients are (and must be) constant in time $t$,  as standard in the literature, due to the preservation of the symplectic structure. Because of this,  we can simply evaluate these formulas at late times, $t>t_{\rm off}$, where the explicit form of $\phi^{\rm out}_{\omega\ell m}$ is known in (\ref{outlate}).  The remaining  task is then to determine the time evolution of the in modes $\phi^{\rm in}_{\omega\ell m}$ by solving (\ref{ode}), with initial data (\ref{inearly}) and with the time-dependent boundary condition induced by the oscillating stellar surface. 

The Bogoliubov coefficients are not arbitrary, but must satisfy the following unitarity conditions,
\bea
\sum_{\vec k''}  (\alpha_{\vec k \vec k''}\overline{\alpha_{\vec k' \vec k''}}- \beta_{\vec k \vec k''}\overline{\beta_{\vec k' \vec k''}})&=&\delta_{\vec k,\vec k'}\, ,\label{constraint1}\\
\sum_{\vec k''}  (\alpha_{\vec k \vec k''}{\beta_{\vec k' \vec k''}}- \beta_{\vec k \vec k''}{\alpha_{\vec k' \vec k''}}) &=&0\, .\label{constraint2}
\eea
These relations ensure the unitarity of the Bogoliubov transformation between the in and out Fock representations, ultimately rooted in the time evolution of the field modes.

\subsection{Orthogonality and normalization of the field modes}
\label{sec:orthogonality}

As mentioned above, the normalization factor $A_{\omega\ell}$ is obtained from the Klein-Gordon inner product by imposing $\langle \phi^{\rm in}_{\omega\ell m}, \phi^{\rm in}_{\omega'\ell' m'}\rangle_+=\delta(\omega-\omega')\delta_{\ell \ell'}\delta_{m m'}$. Since the symplectic structure is independent of the choice of hypersurface $\Sigma_t$, the inner product may be evaluated at early times, $t<t_{\rm on}$, where the in-basis of field modes takes the particularly simple form in (\ref{inearly}):
\bea
\langle\phi^{\rm in}_{\omega\ell m},\phi^{\rm in}_{\omega'\ell' m'}\rangle_+&=&\frac{i}{\hbar} \int_{\Sigma_t} \,d\Sigma_t  \, t^a\, (\overline{\phi^{\rm in}_{\omega\ell m}} \, \nabla_a \phi^{\rm in}_{\omega'\ell' m'}- \phi^{\rm in}_{\omega'\ell' m'} \,\nabla_a \overline{\phi^{\rm in}_{\omega\ell m}}) \nonumber\\
&=&\frac{i}{\hbar}  \int_{R_0}^{\infty} dr r^2\left(1-\frac{2M}{r}\right)^{-1} d\cos\theta\, d\phi\,   (\overline{\phi^{\rm in}_{\omega\ell m}} \, \partial_t \phi^{\rm in}_{\omega'\ell' m'}- \phi^{\rm in}_{\omega'\ell' m'} \,\partial_t \overline{\phi^{\rm in}_{\omega\ell m}})\nonumber\\
&=&\overline{A_{\omega\ell }}A_{\omega'\ell } (\omega+\omega')e^{-i(\omega'-\omega)t} \delta_{\ell \ell'}\delta_{mm'}\int_{R_0}^{\infty} dr r^2\left(1-\frac{2M}{r}\right)^{-1}  \overline{R_{\omega\ell}}  R_{\omega'\ell}\, .
 \label{kginner}
\eea

In practical numerical implementations, however, the radial integration must be truncated at some finite cutoff $r_{\rm max} < \infty$. This truncation is not innocuous: as a result of it,  a suitable choice of boundary conditions for the radial functions $R_{\omega\ell}(r)$ at the fiducial value $r = r_{\rm max}$ must be imposed  in order to preserve  the orthogonality of the basis modes necessary for the quantization. To make this explicit, let us rewrite Eq.~(\ref{ode2}) in the standard Sturm-Liouville form,
\bea
\left[\frac{d}{dr}\left(p(r)\frac{d}{dr}\right)+\omega^2 \rho(r)-q_{\ell}(r)\right]R_{\omega\ell}(r)=0\, , \label{sl0}
\eea
where  $p(r)=r^2f(r)$, $\rho(r)=\frac{r^2}{f(r)}$, $q_{\ell}(r)=\ell(\ell+1)$. This naturally induces the  inner product
\bea
\langle R_{\omega\ell}, R_{\omega'\ell}\rangle_{\rm SL}:=\int_{R_0}^{r_{\rm max}}dr \rho(r)R_{\omega\ell}(r)R_{\omega'\ell}(r)\, .\label{innerSL}
\eea
Then, for $\omega\neq \omega'$  one readily finds
\bea\label{slproblem}
(\omega^2-\omega'^2)\langle R_{\omega\ell}, R_{\omega'\ell}\rangle_{\rm SL}&:=& \int_{R_0}^{r_{\rm max}}dr \left[- R_{\omega\ell}\frac{d}{dr}\left(p(r)\frac{d}{dr}\right)R_{\omega'\ell} +R_{\omega'\ell}\frac{d}{dr}\left(p(r)\frac{d}{dr}\right)R_{\omega\ell}  \right]\nonumber\\
&=& \left. p(r) R_{\omega'\ell}\partial_r R_{\omega\ell}\right|_{R_0}^{r_{\rm max}}- \left. p(r) R_{\omega\ell}\partial_r R_{\omega'\ell}\right|_{R_0}^{r_{\rm max}}\nonumber\\
&=& p(r_{\rm max})\left[ R_{\omega'\ell}(r_{\rm max}) R'_{\omega\ell}(r_{\rm max}) - R_{\omega\ell}(r_{\rm max}) R'_{\omega'\ell}(r_{\rm max})\right]\,, 
\eea
where in the last step we used the early-time boundary condition $R_{\omega\ell}(t,R_0)=0$. Equation (\ref{slproblem}) shows that the orthogonality of modes with different $\omega$ depends crucially on the fiducial boundary conditions imposed at $r=r_{\rm max}$. In particular, this result indicates that naively matching the radial functions to their Minkowski counterparts at the outer boundary, $R_{\omega\ell}(r_{\rm max}) = j_\ell(\omega r_{\rm max})$, does not in general yield an orthogonal set of modes, even if $r_{\rm max}>>M$. Consequently, it is important to choose suitable boundary conditions at  $ r_{\rm max}<\infty$, so as to correctly reproduce both orthogonality and the Minkowski behavior in the $r_{\rm max}\to \infty$ limit.

Our strategy here will be to interpret the fiducial value $r_{\rm max}$  as a numerical representation of spatial infinity, which has been effectively brought to a finite coordinate distance. Since the radial Minkowski modes tend to zero  at large $r$, in order to mimic this asymptotic behavior within a finite computational domain while preserving orthogonality of the modes, we impose that $r_{\rm max}$ coincides with a node of the radial functions $R_{\omega\ell}(r)$ for each $(\omega,\ell)$. This prescription is the analogue of placing the system inside a large spherical box and taking the limit $r_{\rm max}\to\infty$. In the latter limit,  this prescription reproduces the asymptotically flat behavior and the corresponding continuum normalization of the physical problem.

To gain intuition, it is useful to consider the analogous problem in Minkowski spacetime. In this case, the quantum field admits the standard expansion
\bea\label{auxiliary}
\hat \phi(x) =\sqrt{\hbar} \int_{0}^{\infty}d \omega\sum_{\ell=0}^{\infty}\sum_{m=-\ell}^{+\ell} a_{\omega\ell m} A_{\omega\ell} e^{-i \omega t}j_{\ell}( \omega r)Y_{\ell m} (\theta,\phi) + h.c. \, .
\eea
One way to recover this expression is to impose the boundary conditions $R_{\omega\ell}(0) = \delta_{\ell 0}$ and $R_{\omega\ell}(r_{\rm max}) = 0$
on a fiducial bounded interval $[0,r_{\rm max}]$, with $r_{\rm max}$ sufficiently large, and subsequently take the limit $r_{\rm max} \to \infty$.
In Minkowski spacetime the general radial solution of (\ref{ode2}) is a linear combination of $j_\ell(\omega r)$ and $y_\ell(\omega r)$. Regularity at the origin excludes $y_\ell$, while the outer boundary condition implies $j_\ell(\omega r_{\rm max}) \simeq \sin(\omega r_{\rm max}) = 0$, leading to the discrete spectrum $\omega_n \simeq n\pi / r_{\rm max}$ with $n \in \mathbb{N}$, for sufficiently high $r_{\max}$. Thus, the modes $\omega$ are discretized and the quantum field takes the form
\bea
\hat \phi(x) =\frac{\sqrt{\hbar}\,\pi}{r_{\max}} \sum_{n=0}^\infty\sum_{\ell=0}^{\infty}\sum_{m=-\ell}^{+\ell} a_{\omega_n\ell m} A_{\omega_n\ell}  e^{-i \omega_n t} j_{\ell}( \omega_n r)Y_{\ell m} (\theta,\phi) + h.c. \, ,
\eea
where the factor $r_{\rm max}^{-1}$ ensures that $A_{\omega_{n}\ell}$ preserves the  dimensions of ${\rm length}^{-1/2}$. Taking the limit $r_{\rm max} \to \infty$, the discrete sum over $\omega_n$ turns into the continuous integral (\ref{auxiliary}) via
\bea
\lim _{L \rightarrow \infty} \frac{\pi}{L} \sum_{n \in \mathbb N} f\left(\frac{\pi n}{L}\right)= \int_{0}^{+\infty} d \omega f(\omega)\, .
\eea 

Motivated by this construction, we adopt an analogous strategy in the curved spacetime exterior to the star. Instead of imposing $R_{\omega\ell}(r_{\rm max}) = j_\ell(\omega r_{\rm max})$, which does not lead to an orthogonal set of basis modes in (\ref{kginner}), we impose the fiducial Dirichlet  condition, $R_{\omega\ell}(r_{\rm max}) = 0$, for a sufficiently large cutoff $r_{\rm max} \gg M$. Such vanishing value at $r=r_{\rm max}>>M$ is expected to approximate reasonably well the fall-off of the Minkowski modes at spatial infinity, $\lim_{r\to \infty}j_{\ell}(\omega r)=0$.
More importantly, equation (\ref{slproblem})  implies
\bea
\langle R_{\omega\ell}, R_{\omega'\ell} \rangle_{\rm SL} \propto \delta_{\omega \omega'} ,
\eea
yielding a complete orthonormal set of radial modes. The normalization factor is then given by
\bea\label{Akl-squared-early-times}
|A_{\omega\ell}|^2= \frac{r_{\max}}{2\pi \omega}\left( \int_{R_0}^{r_{\rm max}} dr r^2\left(1-\frac{2M}{r}\right)^{-1}   R_{\omega\ell }^2(r) \right)^{-1}\, ,
\eea
 obtained from (\ref{kginner})  after demanding the orthonormality condition $\langle \phi_{\omega\ell m},\phi_{\omega'\ell' m'}\rangle_+=\frac{r_{\max}}{\pi}\delta_{\omega,\omega'}\delta_{\ell,\ell'}\delta_{m,m'}$.

\subsection{Modeling stellar oscillations and comoving coordinates}

The oscillatory motion of the stellar surface $R_{0}(t)$ induces a time-dependent boundary condition in the field equation (\ref{ode}). Solving partial differential equations with moving boundaries is  technically challenging in practice, especially in contexts where mode orthogonality and  conservation of the symplectic structure must be maintained throughout the evolution.  To overcome this difficulty,  we  introduce a suitable coordinate transformation that maps this time-dependent boundary to a fixed coordinate location. This strategy is conceptually analogous to the use of conformal transformations in 1+1 dimensions to treat the Klein-Gordon equation with time-dependent boundaries, as originally discussed in~\cite{10.1063/1.1665432}

\begin{figure}[!htbp]
  \centering  \includegraphics[width=0.6\columnwidth]{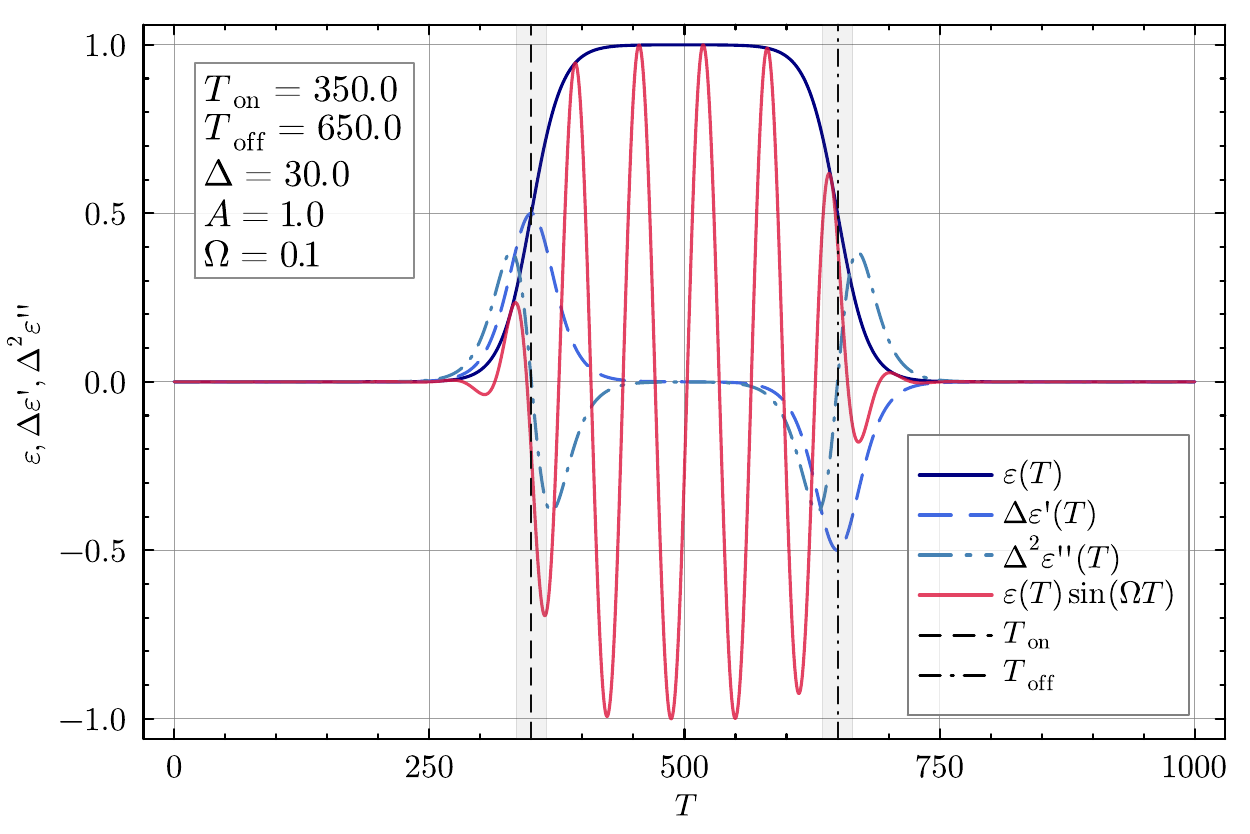}
  \caption{Switching function $\epsilon(T)$ for some example parameters, its first and second derivatives and the resulting oscillations. The left and right gray bands centered around $T_{\text{on}}$ and $T_{\text{off}}$ have width $\Delta$.}
  \label{fig:epsilon-display}
\end{figure}

We assume that the radial oscillations of the stellar surface can be modeled as  \cite{PhysRevLett.12.114,1964ApJ...140..417C,Kokkotas2001,PhysRevLett.127.191101}
\bea
R_{0}(t) &=& {R_{0}} + M\epsilon(t)\sin(\Omega t)\, , \label{stellaroscillation}
\eea
where $\Omega \in \mathbb{R}$ represents the oscillation frequency of the star, $R_0$ is the initial and final stellar radius, and $\epsilon(t)$ is a dimensionless switching function, which vanishes  at early and late times when the star is static. Both $\Omega$ and $\epsilon(t)$ are treated as  fixed, background parameters, determined by internal physics of the star. For definiteness, we adopt a smooth switching function of the form 
\bea
    \varepsilon(t; t_{\text{on}}, t_{\text{off}}, \Delta, A) = \frac{A}{4} \left[1 + \tanh\left(\frac{t - t_{\text{on}}}{\Delta}\right)\right] \left[1 - \tanh\left(\frac{t - t_{\text{off}}}{\Delta}\right)\right]\, , \label{eq:windowing-function}
\eea
whose  profile is shown in Fig. \ref{fig:epsilon-display}. This function decays exponentially in time as $\epsilon(t)\sim A\, e^{-2t/\Delta}$, showing that the parameter $\Delta$ controls the characteristic timescale over which the oscillations are switched  off. In this effective description, $\Delta$ can be interpreted as setting an imaginary part of an \emph{effective} quasi-normal mode frequency associated with the stellar oscillations, with real part  given by $\Omega$. Accordingly, the pair $(\Omega,\Delta)$ can be regarded as phenomenological parameters encoding the oscillatory response of the star and entirely determined by its microphysics, whereas the amplitude $A$ depends on the initial perturbation that triggers the oscillations.

We note that, within our framework,  the amplitude $A$ can in principle be chosen arbitrarily large, as long as the stellar surface remains outside the Schwarzschild radius at all times, $R_0 - M A > 2M$. Likewise, the parameters $\Delta^{-1}$ and $\Omega$, which control the characteristic velocity and acceleration scales of the motion, are not restricted to small values. No slow-motion, small-amplitude, or weak-field expansions are employed in our analysis, and the numerical scheme described below is designed to handle this fully relativistic regime.

We now perform the coordinate transformation $(t,r,\theta,\phi) \to (T,z,\theta,\phi)$ defined by
\bea
T(t,r) &=& t, \label{eq:coordT}\, , \\
z(t,r) &=& r - M\,\epsilon(t)\,\sin(\Omega t).
\label{eq:coordiante-change}
\eea
Under this transformation, the original time-dependent boundary condition $R_\ell(t,R_0(t))=0$ is mapped to the fixed Dirichlet condition 
\bea
R_\ell(T,R_0)=0\, , \label{innerboundary}
\eea
corresponding to a rigid boundary located at the constant spatial position $z=R_0$. In these coordinates, points on the stellar surface remain at fixed spatial coordinates and are therefore comoving with the star, even though they correspond to non-static worldlines in the original Schwarzschild coordinates. We will henceforth refer to $(T,z,\theta,\phi)$ as \textit{comoving coordinates}.

The next step now is to reformulate both the  Schwarzschild line element \eqref{metric} as well as the field equation \eqref{ode} in this new set of coordinates. Using the \texttt{xAct} package \cite{xAct} of \textit{Mathematica}, the components of the Schwarzschild metric  take the form
\bea
ds^2= - \left[f(r(T, z))- \frac{\dot{S}^2(T)}{f(r(T, z) ) }\right]dT^2+2 \frac{\dot{S}(T)}{f(r(T, z) )} dTdz+ \frac{1}{f(r(T, z))}dz^2+r^2(T, z)d\Omega^2\, , \label{schwoscillating}
\eea
where  $r(T, z) \equiv  z+ M\epsilon(T) \sin (T \Omega)$, ${S}(T) = M\epsilon(T) \sin(\Omega T)$ and overdots denote derivatives with respect to $T$. On the other hand, the Klein-Gordon equation (\ref{ode}) in comoving coordinates \eqref{eq:coordiante-change} for the ansatz $\phi_{\omega \ell m} (x) =  \sqrt{\hbar}\, A_{\omega\ell} R_{\omega\ell}(T, z) Y_{\ell m}(\theta, \phi)$ becomes
  \bea
    \left[ - \frac{\partial^2}{\partial T^2} + 2\dot{S}  \frac{\partial^2}{\partial T \partial z} +  \left[ f^{2} - \dot{S}^2  \right] \frac{\partial^2}{\partial z^2}  + \left[ \frac{2M f}{r^2(T,z)} + \frac{2 f^{2}}{r(T,z)} +\ddot{S}(T)  \right] \frac{\partial}{\partial z} - \frac{\ell (\ell+1) f}{r^2(T,z)}\right] R_{\omega\ell}(T, z)= 0\, . \label{eq:ReggeWheeler-SC-oscillating-coords}
  \eea
  This equation correctly reduces to the standard Regge-Wheeler equation in Schwarzschild spacetime (\ref{ode}) in the limit $\Omega \to 0$, as well as at sufficiently early and late times for which $\epsilon(T) \to 0$. Likewise, taking the limit $M \to 0$ recovers the Klein-Gordon equation in Minkowski spacetime expressed in spherical coordinates.

To integrate this PDE on a finite numerical grid, we must again prescribe a boundary condition at a fiducial outer point $z=z_{\rm max}$. This condition should be compatible with the asymptotically flat character of the spacetime and with the orthogonality properties of the initial Sturm-Liouville modes. Since, at sufficiently large radius, the effects of the stellar oscillations are expected to be negligible and the field propagates freely to leading order, we impose that the modes evolve at $z_{\rm max}$ according to their asymptotic free time dependence,
\bea
R_{\omega\ell} (T , z_{\rm max}) = e^{-i\omega T}R_{\omega\ell}(z_{\rm max} + S(T)), \quad \forall T\in \mathbb R\, , \label{freeBC}
\eea
where $R_{\omega\ell}(z)$ denotes the corresponding eigenfunction of the static Sturm-Liouville problem (\ref{sl0}). In the original $(t,r)$ coordinates, this condition is equivalent to requiring $R_{\omega\ell} (t,r_{\rm max})=e^{-i\omega t}R_{\omega \ell}(r_{\rm max})$ for any $t$. At early and late times this equation reduces to the Dirichlet condition $R_{\omega\ell} (t,r_{\rm max})=0$ introduced in the previous subsection.

Finally, the Klein-Gordon  inner product in comoving coordinates takes the form
\bea
  \langle \phi_{\omega \ell m}, \phi_{\omega' \ell' m'} \rangle_+   =  i \delta_{\ell \ell'} \delta_{m m'} \overline{A_{\omega\ell }}A_{\omega'\ell } \int_{R_0}^{r_{\rm max}-S(T)} d z \frac{[z + S(T)]^2}{f(r(T,z))} \left[ \overline{R_{\omega \ell}} \left(  \partial_T R_{\omega' \ell'} + \dot S \partial_z R_{\omega' \ell'} \right) -R_{\omega' \ell'} \left(\partial_T \overline{R_{\omega \ell}} + \dot S \partial_z \overline{R_{\omega \ell}} \right)  \right] \, .\label{eq:inner-prod-oscillating-coords}
\eea
The spatial integral is truncated to $z_{\rm max}(T)= r_{\rm max}-S(T)$ to ensure the orthogonality of the field modes for $\omega'\neq \omega$ when implementing this formula in the numerical code with the fiducial boundary condition (\ref{freeBC}). This orthogonality  is easily seen at early times when $S(T)=0$,  and since  the symplectic structure is preserved in time thanks to (\ref{freeBC}), we can safely claim $\langle \phi_{\omega \ell m}, \phi_{\omega' \ell' m'} \rangle_+\propto \delta_{\omega,\omega'}$.  Demanding now the condition $\langle \phi_{\omega\ell m},\phi_{\omega'\ell' m'}\rangle_+=\frac{r_{\max}}{\pi}\delta_{\omega,\omega'}\delta_{\ell,\ell'}\delta_{m,m'}$, we can solve for $A_{\omega\ell}$, yielding 
\bea
A_{\omega\ell}   =  \sqrt{\frac{r_{\rm max}}{i\pi}}  \left[\int_{R_0}^{r_{\rm max}-S(T)} d z \frac{[z + S(T)]^2}{f(r(T,z))} \left[ \overline{R_{\omega \ell}} \left(  \partial_T R_{\omega' \ell'} + \dot S \partial_z R_{\omega' \ell'} \right) -R_{\omega' \ell'} \left(\partial_T \overline{R_{\omega \ell}} + \dot S \partial_z \overline{R_{\omega \ell}} \right)  \right] \right]^{-1/2}\, ,\label{norm}
\eea
at any instant of time $T$.

\section{Numerical methods}
\label{numerical_methods}

In this section we describe the methodology employed to solve the time evolution of the field modes and for computing the Bogoliubov coefficients.

\subsection{Time evolution of the radial modes\label{sec:numerical-methods-resonant}}

To solve the field equation~\eqref{eq:ReggeWheeler-SC-oscillating-coords} for $R_{\omega\ell}(T,z)$, we employ the Method of Lines (MoL). The spatial coordinate
$z \in [z_0, r_\text{max}]$ is discretized into a finite grid $\{z_i\}_{i=0}^{N}$, with $z_N = r_\text{max} = z_{\rm max}(0)$, and spatial derivatives are approximated using finite-difference schemes. This procedure reduces the PDE to a system of coupled ODEs for the discrete variables $R_{\omega\ell}(T, z_i)$. We adopt a uniform spatial discretization of the form $z_i = z_0 + (z_N - z_0)\,\frac{i}{N}$, for $i = 0, \ldots, N$.

The Dirichlet condition (\ref{innerboundary}) and the ``free-evolution'' condition (\ref{freeBC})  are imposed  at both ends of the numerical grid, $z = z_0$ and $z = z_N$, respectively. To mitigate numerical artifacts associated with the presence of a rigid outer boundary in our grid, the computational domain is further extended to $z \in [z_0, 2 z_N]$ for convenience. 
Since the propagation speed of the field modes is $c = 1$, any spurious radiation generated at the artificial boundary $z = 2 z_N$ can only reach the physical domain, whose outer boundary is at $z = z_{\rm max}(T)$, for evolution times $T \gtrsim r_{\rm max} + T_\text{on}$. Accordingly, all simulations presented in this work are restricted to evolution times satisfying $T < r_{\rm max}$. This construction ensures that the physical domain $[z_0, z_{\rm max}(T)]$ remains unaffected by boundary-induced artifacts throughout the evolution in the grid. 

The resulting system of ODEs is integrated in time using Verner's ``Most Efficient'' 9/8 Runge-Kutta method (Verner9) \cite{Verner2010}. This scheme combines high-order accuracy with adaptive step-size control. The numerical implementation is carried out using the \texttt{DifferentialEquations.jl} library in \texttt{Julia} \cite{Rackauckas2017}.

Once a solution $R_{\omega\ell}(T,z)$ for the radial modes has been obtained, we normalize it via $\hat R_{\omega\ell}(T,z):=A_{\omega \ell}R_{\omega\ell}(T,z)$ according to formula (\ref{norm}). The radial integral is computed using a composite  Simpson's $1/3-3/8$ rule. Then, to assess the robustness and consistency of the numerical framework we compute the norm squared $\langle \phi_{\omega \ell m}, \phi_{\omega \ell m} \rangle_+$ of each mode $\phi_{\omega \ell m}$, not simply at early times using a truncated version of Eq.~\eqref{kginner}, but  at each instant of time via the Klein-Gordon inner product in comoving coordinates, Eq.~\eqref{eq:inner-prod-oscillating-coords}, and verify that it remains constant throughout the entire evolution within numerical accuracy. This provides a direct numerical check of the time-independence of the symplectic structure and of the overall numerical framework.  Numerical deviations from perfect preservation set a criterium for the reliability of the results.

\subsection{Determination of the Cauchy data}

Initial conditions for the  equation~\eqref{eq:ReggeWheeler-SC-oscillating-coords} for the in-modes are imposed during the stationary regime prior to the onset of stellar oscillations, such that the field modes are of positive frequency with respect to the early-time Killing field $\partial_T$. This choice defines the in-vacuum state. Specifically, at some initial time $T=T_0<T_{\rm on}$ we impose
\bea
R^{\rm in}_{\omega\ell}(T_0, z)&=&e^{-i\omega T_0}R_{\omega\ell}(z)\, , \label{in1}\\
 \partial_T R^{\rm in}_{\omega\ell}(T_0, z)&=&-i\omega e^{-i\omega T_0}R_{\omega\ell}(z)\, . \label{in2}
\eea
Similarly, for the out-modes we impose, at some final time $T=T_{\rm max}>T_{\rm off}$,
\bea
R^{\rm out}_{\omega\ell}(T_{\rm max}, z)&=&e^{-i\omega T_{\rm max}}R_{\omega\ell}(z)\, , \label{out1}\\
 \partial_T R^{\rm out}_{\omega\ell}(T_{\rm max}, z)&=&-i\omega e^{-i\omega T_{\rm max}}R_{\omega\ell}(z)\, . \label{out2}
\eea
To determine the radial profile $R_{\omega\ell}(z)$ in both cases, we subsitutte the ansatz $R_{\ell}(T,z)\sim e^{-i \omega T}R_{\omega\ell}(z)$ in  \eqref{eq:ReggeWheeler-SC-oscillating-coords} in the non-oscillating limit $\Omega \to 0$, yielding (\ref{ode2}). After multiplying by $z^2/f(z)$ and rearranging  terms, we  can recast the problem into a standard Sturm-Liouville form, 
\bea
  \left[ \frac{d}{dz} \left( p(z) \frac{d}{dz} \right) + \omega^{2} \rho(z) - q_{\ell}(z) \right] R_{k \ell}(z) = 0\, , \label{sl}
\eea
with positive eigenvalues $\omega^{2}$, coefficient functions $p(z) = z^{2}f(z)$, $q(z) = \ell (\ell + 1)$, and weight function $\rho(z) = z^{2}/f(z)$. For each fixed $\ell$, Sturm-Liouville theory guarantees that, subject to Dirichlet boundary conditions $R_{\omega \ell}(z_0)=R_{\omega \ell}(z_{N})=0$, there will be an infinite,  discrete set of real eigenvalues $\omega^2$ and a corresponding complete, orthogonal set of eigenfunctions $R_{\omega\ell}(z)$  with respect to the inner product $\langle R_{\omega\ell}, R_{\omega'\ell}\rangle_{SL}=\int_{z_0}^{z_N}dz\rho(z)R_{\omega\ell}(z)R_{\omega'\ell}(z)$ \cite{SturmLiouvilleNumericalSolutions}. For each $\ell$, we will label each of these eigenvalues by $\omega_{n\ell}$, with $n\in \mathbb N$, and denote the associated eigenfunctions by $R_{\omega_{n\ell}\ell}(z) \equiv R_{n\ell}(z)$.

The Sturm-Liouville eigenvalue problem is solved numerically using a shooting method for boundary-value problems (see e.g. \cite{SturmLiouvilleNumericalSolutions, Press2007}) and employing finite-difference methods, yielding initial estimates for the eigenvalues $\omega_{n\ell}^2$ and eigenfunctions $R_{n\ell}(z)$. The shooting method is a classical technique for solving boundary-value and eigenvalue problems. The differential equation is first recast as an initial-value problem by guessing the unknown eigenvalue and integrating from one boundary. The guessed parameter is then iteratively adjusted until the solution satisfies the required boundary condition at the opposite endpoint. These solutions are subsequently refined using the Verner9 integration scheme to achieve machine-precision accuracy on the chosen grid. 

\subsection{Computation of the Bogoliubov coefficients}

Once the normalized radial modes $\hat R^{\rm in}_{n\ell}(T,z)$ and $\hat R^{\rm out}_{n\ell}(T,z)$ have been obtained for all $T\in [T_0,T_{\rm max}]$, we proceed to compute the Bogoliubov coefficients. The $\beta$ coefficient can be  evaluated on any constant-$T$ hypersurface as
\bea
  \beta_{\vec {k}\vec {k}'} &=&\frac{\Omega(\phi^{\rm out}_{\vec k'},  \phi^{\rm in}_{\vec k})}{\Omega(\phi^{\rm out}_{\vec k},  \overline{\phi^{\rm out}_{\vec k}})}=(-1)^m\delta_{\ell, \ell'} \delta_{m, -m'} \beta_{n\ell,n'\ell}\, ,
  \eea
  where $\vec k$ here stands for $(n,\ell,m)$, and 
  \bea
  \beta_{n\ell,n'\ell} &=&  \int_{z_0}^{r_{\rm max}-S(T)} dz  \frac{[z + S(T)]^2}{f(r(T,z))}    \left[ \hat R^{\rm in}_{n \ell} \left( \partial_T {\hat R^{\rm out}_{n' \ell}} + \dot S \partial_z {\hat R^{\rm out}_{n' \ell}} \right) -  {\hat R^{\rm out}_{n' \ell}} \left(  \partial_T \hat R^{\rm in}_{n \ell} + \dot S \partial_z \hat R^{\rm in}_{n \ell} \right)  \right]\, . \label{betacoefficients}
  \eea
The Kronecker deltas arise as a consequence of the spherical symmetry of the background, and indicate that particle pairs are produced with the same orbital angular-momentum  number $\ell$ and opposite magnetic number, $m'=-m$ (the factor $(-1)^m$ follows from the identity $Y_{\ell,-m}=(-1)^m \overline{Y_{\ell m}}$). As for the norm $\langle \phi_{\omega \ell m}, \phi_{\omega \ell m} \rangle_+$, the symplectic structure is conserved in time, and therefore the Bogoliubov coefficients can in principle be evaluated on any hypersurface. At sufficiently late times, for instance, the expression above reduces to a simpler equation using (\ref{outlate})
\bea
\beta_{n\ell,n'\ell}= \int_{R_0}^{r_{\rm max}} dz \,z^2 \left(1-\frac{2M}{z}\right)^{-1} \left[\hat R^{\rm in}_{n \ell} \,\omega_{n'\ell} -i    \partial_T \hat R^{\rm in}_{n \ell}   \right]{\hat R^{\rm out}_{n' \ell}} \, .
\eea
Similarly, the $\alpha$ coefficients are obtained as
\bea
\alpha_{\vec k\vec k'}=\frac{\Omega(\overline{\phi^{\rm out}_{\vec k'}},  \phi^{\rm in}_{\vec k})}{\Omega(\overline{\phi^{\rm out}_{\vec k}},  \phi^{\rm out}_{\vec k})}=\delta_{\ell, \ell'} \delta_{m, m'} \alpha_{n\ell,n'\ell}\, , 
\eea
with 
\bea
\alpha_{n\ell,n'\ell}  &=& -i\int_{z_0}^{r_{\rm max}-S(T)} dz  \frac{[z + S(T)]^2}{f(r(T,z))}    \left[\hat R^{\rm in}_{n \ell} \left( \partial_T \overline{\hat R^{\rm out}_{n' \ell}} + \dot S \partial_z \overline{\hat R^{\rm out}_{n' \ell}} \right) -  \overline{\hat R^{\rm out}_{n' \ell}} \left(  \partial_T \hat R^{\rm in}_{n \ell} + \dot S \partial_z \hat R^{\rm in}_{n \ell} \right)  \right]  \label{alphacoefficients}\\
&=&\int_{R_0}^{r_{\rm max}} dz \,z^2 \left(1-\frac{2M}{z}\right)^{-1} \left[ \hat R^{\rm in}_{n \ell} \,\omega_{n'\ell} +i    \partial_T \hat R^{\rm in}_{n \ell}   \right]\overline{\hat R^{\rm out}_{n' \ell}} \, , \nonumber
\eea
where the last equality is evaluated at late times using (\ref{outlate}), for completeness. In general, neither $\alpha_{n\ell,n'\ell}$ nor $\beta_{n\ell,n'\ell}$ are proportional to $\delta_{nn'}$, reflecting the lack of spatial homogeneity in the background (i.e. absence of radial translational symmetry).

Although it  suffices to evaluate $\alpha_{n\ell,n'\ell}$  and $\beta_{n\ell,n'\ell}$  at late times, we perform the calculation for all  values of $T$ within the integration domain using formulas (\ref{alphacoefficients}) and (\ref{betacoefficients}),  respectively, in order to further assess the numerical stability and accuracy of the simulations. Again, the radial integrals are computed numerically using a composite  Simpson's $1/3-3/8$ rule. The residual variations observed in the coefficients when computed at different values of $T$ provide an estimate of the accumulated numerical error and therefore an upper bound on the reliability of the results presented below.

Finally, the unitarity conditions (\ref{constraint1})-(\ref{constraint2}) simplify to 
\bea
C^{1,\ell}_{nn'}\equiv\sum_{n''=1}^{\infty}(\alpha_{n\ell,n''\ell}\overline{\alpha_{n'\ell,n''\ell}}-\beta_{n\ell,n''\ell}\overline{\beta_{n'\ell,n''\ell}})&=&\frac{r_{\rm max}^2}{\pi^2}\delta_{nn'}\, ,\label{constraint1}\\
C^{2,\ell}_{nn'}\equiv\sum_{n''=1}^{\infty} (\alpha_{n\ell,n''\ell}\beta_{n'\ell,n''\ell}-\beta_{n\ell,n''\ell}\alpha_{n'\ell,n''\ell})&=&0\, , \label{constraint2}
\eea
for any $(\ell,m)$. We compute these infinite-dimensional matrices for sufficiently high values of $n,n'$ to verify the reliability of the numerical results for the Bogoliubov coefficients for any given $\ell$. In practice, the infinite sums are truncated at a sufficiently large cutoff $n''_{\rm max}$, and convergence is verified numerically. The factor $r_{\rm max}^2/\pi^2$ arises from the finite-box normalization adopted for the radial modes.

\section{Results}
\label{results}

In this section we  present the numerical results obtained for the oscillating-star toy model. All the simulations have been performed using a \texttt{Julia} implementation of the numerical methods described  in the previous section. The code is publicly available at a \href{https://github.com/ploliver/PCOS}{GitHub repository}, which also provides detailed instructions to reproduce the results presented here.

Throughout this section we fix the stellar compactness $C = 2M/R_0$ to $C = 0.3$ so that $R_0 \simeq 6.7\,M$, corresponding to a typical neutron star with mass $M \sim 1\,M_\odot$ and radius $R \sim 10\,\mathrm{km}$. While this choice provides a representative astrophysical scenario, the numerical framework developed here can be straightforwardly extended to explore more exotic configurations. For instance, some proposals in gravitational-wave astronomy consider ultra-compact objects with compactness as high as $C \sim 1 - {O}(10^{-40})$ \cite{Cardoso_2019}. These extreme gravitational scenarios are expected to enhance quantum effects and therefore the spontaneous particle creation process.

The simulations were carried out on a physical spatial domain defined by $z_0 = \frac{2}{0.3} M$ and $r_{\rm max} = 2100\,M$, discretized using $N = 11200$ uniformly spaced grid points. As discussed in Sec.~\ref{sec:numerical-methods-resonant}, the corresponding extended fiducial domain contains $2N = 22400$ grid points. The system is evolved from an initial time $T_0 = 0$ up to $T_{\rm max} = 2095\,M$.

The parameters defining the switching function~\eqref{eq:windowing-function} are chosen as $A = 1.0$, corresponding to oscillations with an amplitude of $15\%$ of the initial stellar radius, $T_{\rm on} = 765\, M$, $T_{\rm off} = 1330\,M$, and $\Delta = 40\,M$. These values are chosen so that $\epsilon(T_0)$ and $\epsilon(T_\text{max})$ are zero up to machine precision, ensuring consistency with the expected static asymptotic behavior. The stellar oscillation frequency in Eq.~\eqref{stellaroscillation} is set to $\Omega = 0.03\,M^{-1}$, which is of the same order of magnitude as the first allowed momentum values $\omega_{n\ell}$ (see below), allowing us to investigate potential resonant effects in the particle creation process with a relatively small number of modes.

\subsection{Cauchy data}

\begin{figure}[!htbp]
  \centering
  \begin{subfigure}[b]{0.49\textwidth}
     \centering     \includegraphics[width=\columnwidth]{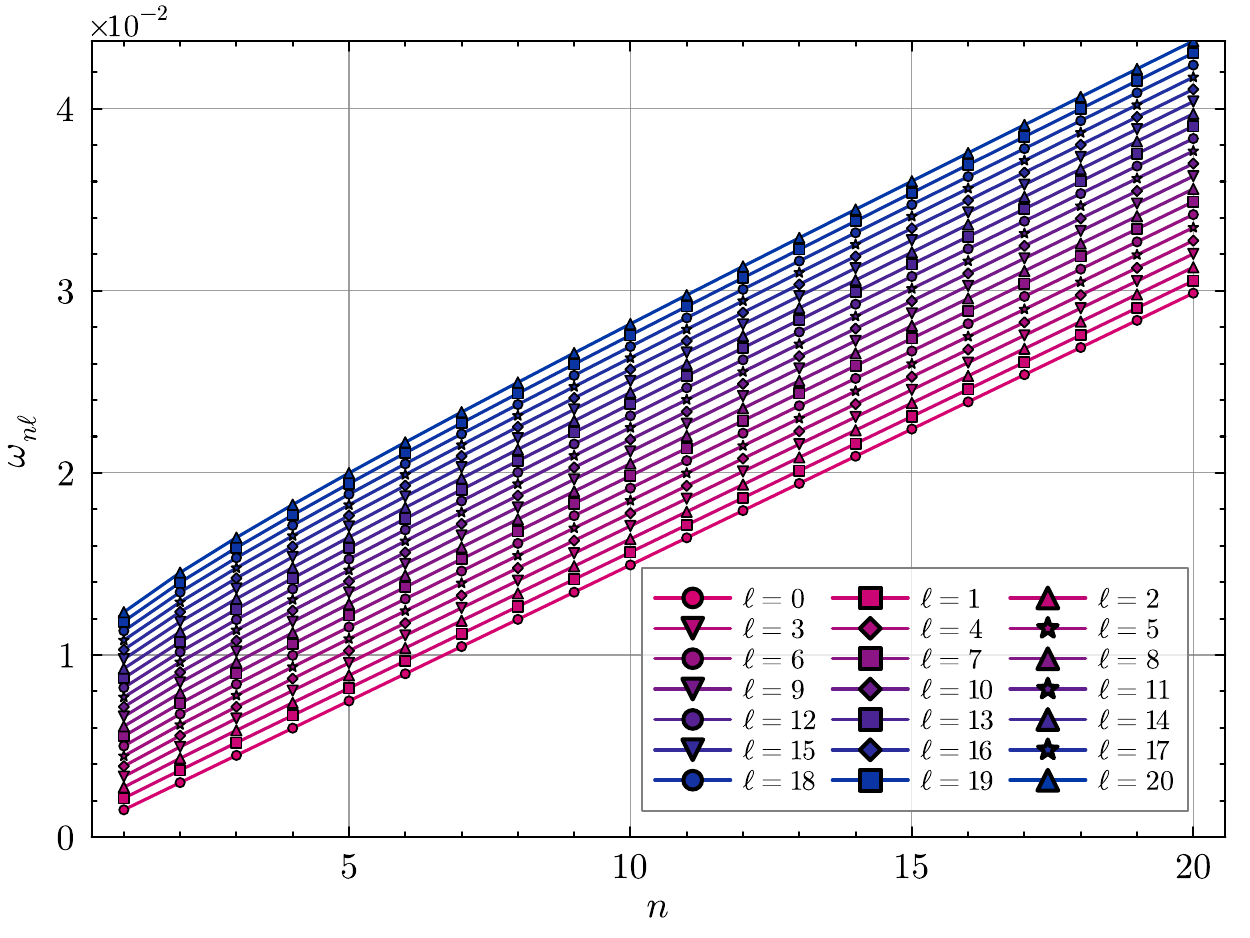}
          \label{fig:sub1}
  \end{subfigure}
  \hfill
  \begin{subfigure}[b]{0.49\textwidth}
     \centering
     \includegraphics[width=\columnwidth]{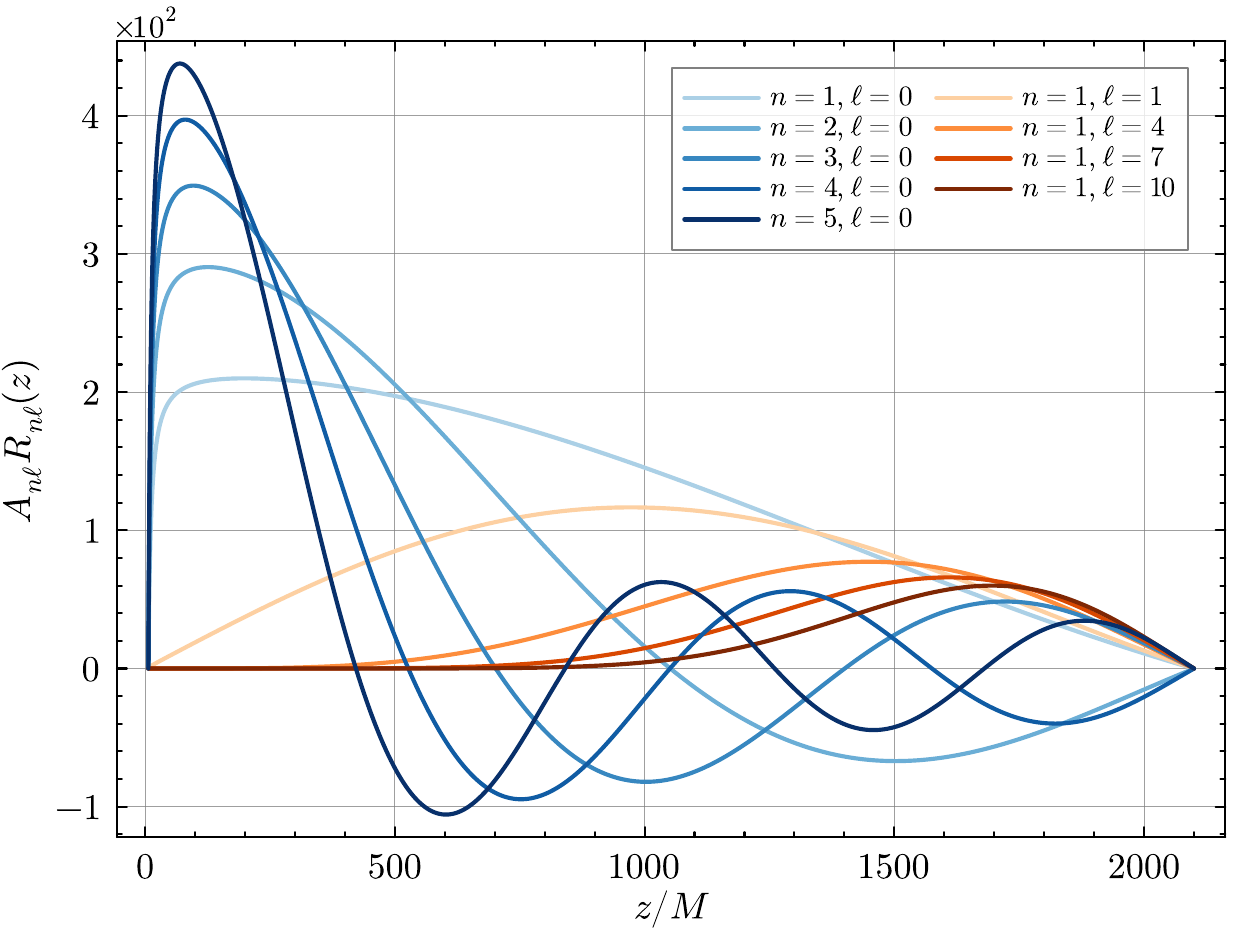}
          \label{fig:sub1}
  \end{subfigure}
  \caption{
  Eigenvalues and eigenfunctions of the Sturm-Liuiville problem in Eq.~\eqref{sl} obtained with boundary conditions $R_{\omega\ell}(z_0)=0$ and $R_{\omega\ell}(z_N)=0$ for $z_0 = 20/3 \,M$ and $z_N = 2100\,M$. Left: first 20 lowest allowed frequencies $\omega_{n\ell}$ for each $\ell \in [0,20]$.  Right: eigenfunctions $R_{n\ell}(z)\equiv R_{\omega_{n\ell}\ell}(z)$  shown for several representative values of $n$ and $\ell$.}
  \label{fig:allowed-momenta-spectrum}
\end{figure}

In Fig.~\ref{fig:allowed-momenta-spectrum} we show the distribution of the lowest allowed momenta $\omega_{n\ell}$ of the Sturm-Liuiville problem in Eq.~\eqref{sl}  with boundary conditions $R_{\omega\ell}(z_0)=0$ and $R_{\omega\ell}(z_{\rm max})=0$, obtained for several values of the angular momentum $\ell$. We observe a monotonic dependence on the radial quantum number $n$, approximately of the form $\omega_{n\ell} \propto n$, as well as the ordering $\omega_{n\ell} > \omega_{n\ell'}$ for $\ell > \ell'$. 

On the other hand, the corresponding eigenfunctions $R_{n\ell}(z)$ for representative values of $n$ and $\ell$ are displayed in Fig.~\ref{fig:allowed-momenta-spectrum}. As expected from Sturm-Liouville theory, the eigenfunction associated with $\omega_{n\ell}$ possesses $n-1$ internal nodes. The shape of the modes also depends on $\ell$. For instance, the lowest modes ($n=1$) for $\ell=0,1,4,7,10$ are all node-free but exhibit distinct radial profiles, as shown in Fig.~\ref{fig:allowed-momenta-spectrum}.

\subsection{Evolved radial modes in time}

As explained above, the computation of the Bogoliubov coefficients requires evolving either the in or the out radial modes in time. The modes $R^{\rm in}_{n\ell}$ are obtained by imposing the initial conditions (\ref{in1})-(\ref{in2}) at $T_0 = 0$ and evolving forward in time, subject to the boundary conditions (\ref{innerboundary}) and (\ref{freeBC}). Similarly, the modes $R^{\rm out}_{n\ell}$ are obtained by imposing the corresponding final conditions (\ref{out1})-(\ref{out2}) at $T_{\rm max} = 1950\,M$ and integrating backward in time toward $T_0 = 0$, with the same boundary prescriptions.

\begin{figure}[!htbp]
  \centering  \includegraphics[width=\columnwidth]{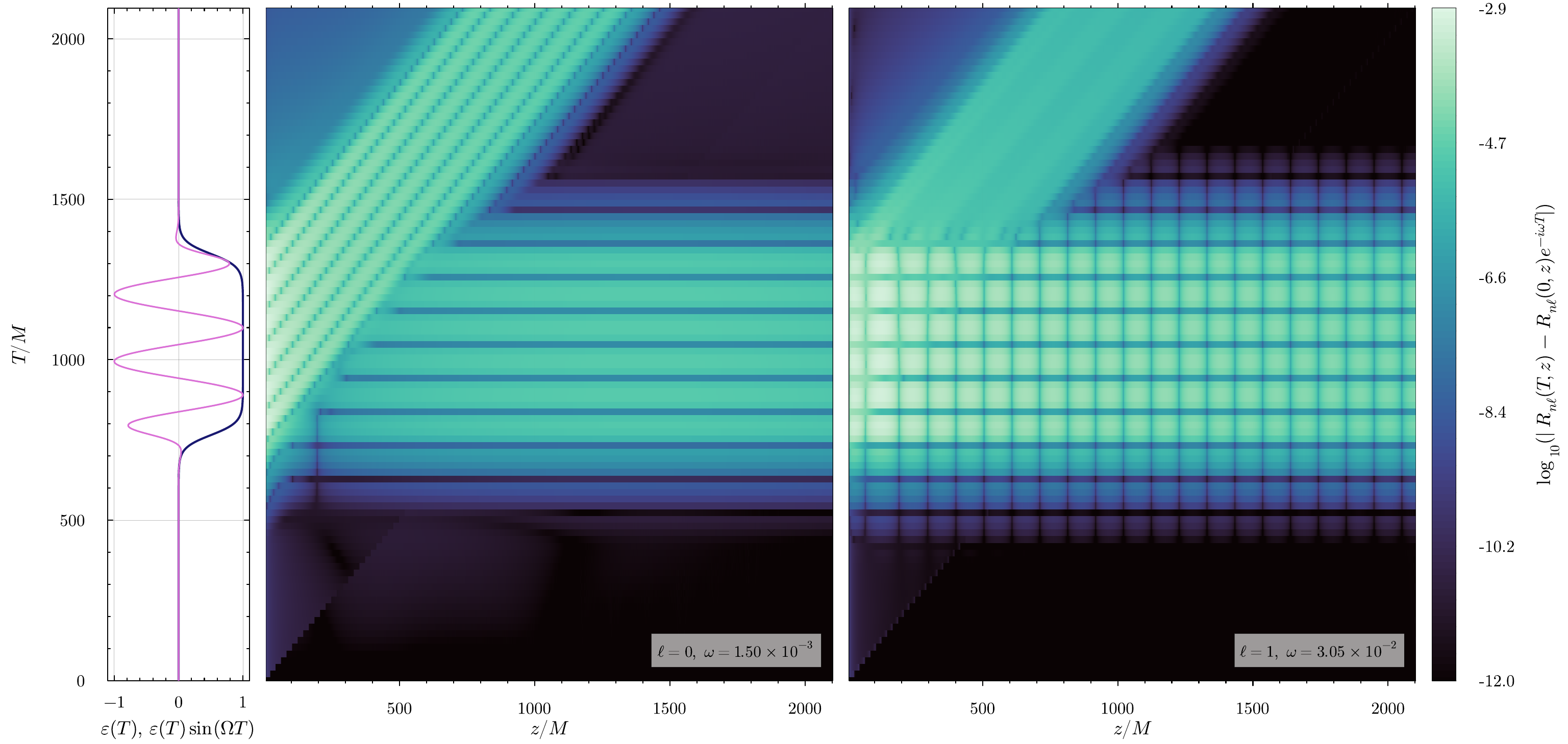}
  \caption{Spacetime diagram in the $(T,z)$ plane showing the deviation of the numerically evolved in radial mode $R^{\rm in}_{n\ell}(T,z)$ from the corresponding freely evolved stationary solution,
$|R^{\rm in}_{n\ell}(T,z)-e^{-i\omega_{n\ell}(T-T_0)}R^{\rm in}_{n\ell}(T_0,z)|$,
displayed in logarithmic scale. 
Left: evolution of the stellar radius, showing the switching function $\varepsilon(T)$ (blue) and the full oscillatory profile $\varepsilon(T)\sin(\Omega T)$ (purple) used in the simulation. 
Center: evolution of the mode $(n,\ell)=(1,0)$, corresponding to the lowest frequency $\omega_{n\ell}$. 
Right: evolution of the mode $(n,\ell)=(20,1)$, whose frequency $\omega_{n1}$ is closest to $\Omega$. 
Both panels exhibit a sequence of outward-propagating wave bursts traveling at the speed of light, originating near the stellar surface at the instants of maximal radial compression and expansion.} 
  \label{fig:time-snapshots-real-only-Schwarzschild}
\end{figure}

Once the stellar oscillations become dynamically significant, their effect on the radial modes can be clearly observed. For instance, Figure~\ref{fig:time-snapshots-real-only-Schwarzschild} displays a two-dimensional spacetime diagram for the  in mode $(n,\ell)=(1,0)$, showing the deviation of $R^{\rm in}_{n\ell}(T,z)$ from the corresponding freely evolved stationary solution. The oscillatory motion of the stellar surface generates outgoing wave-like perturbations superimposed on the free mode. These disturbances originate near the stellar surface at $z_0$ and propagate toward larger values of $z$ at the speed of light, as indicated by their approximately $45^\circ$ inclination in the diagram.

The bursts are produced at those instants of time when the stellar contraction or expansion is most pronounced, corresponding to extrema in the surface acceleration. This behavior is consistent with the expectation that the strongest time-dependent gravitational effects occur when the stellar motion departs most significantly from stationarity. No spurious reflections (``junk'' radiation) are observed within the physical domain during the time interval considered, which confirms the effectiveness of the ``free-evolution'' boundary condition (\ref{freeBC}) at the outer fiducial boundary of the grid.

Similar spacetime diagrams are obtained for other values of $(n,\ell)$. These results illustrate the essential physical mechanism: the oscillatory motion of the star modifies the field modes in a nontrivial way, thereby inducing a change in the vacuum state of the quantum field. Since the perturbations are driven by stellar oscillations with frequency $\Omega$, one anticipates resonant behavior when the relevant mode frequencies satisfy suitable matching conditions involving $\Omega$. This effect will be analyzed quantitatively in Sec.~\ref{sec:calculation-beta}.

With the numerical solutions for $A_{n\ell} R^{\rm in}_{n\ell}(T,z)$ and $A_{n\ell} R^{\rm out}_{n\ell}(T,z)$ at hand, we compute the inner product $\langle \phi^{\rm in}_{\omega\ell m}, \phi^{\rm in}_{\omega\ell m} \rangle$ in comoving coordinates, as given in Eq.~(\ref{eq:inner-prod-oscillating-coords}), for each mode $(n,\ell)$ at every time step $T$ of the simulation. This computation serves as a probe of the accuracy and stability of the simulations. Since the symplectic product (\ref{symp}) is conserved in time, the norm $\langle \phi_{\omega\ell m}, \phi_{\omega\ell m} \rangle^{1/2}$ should remain constant throughout the evolution. 

\begin{figure}[!htbp]
  \centering
  \begin{subfigure}[b]{0.49\textwidth}
     \centering     \includegraphics[width=\columnwidth]{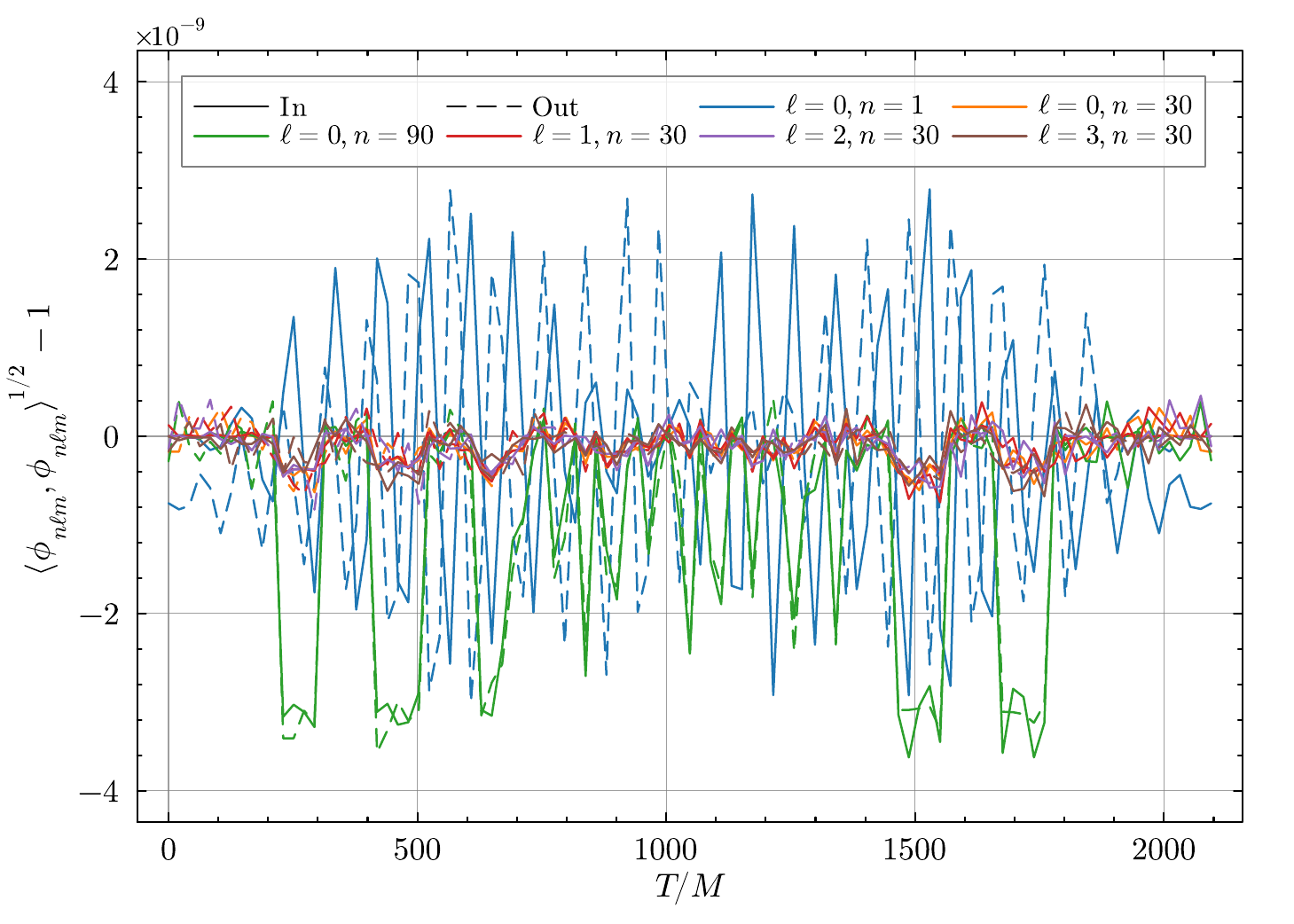}
  \end{subfigure}
  \hfill
  \begin{subfigure}[b]{0.49\textwidth}
     \centering
     \includegraphics[width=\columnwidth]{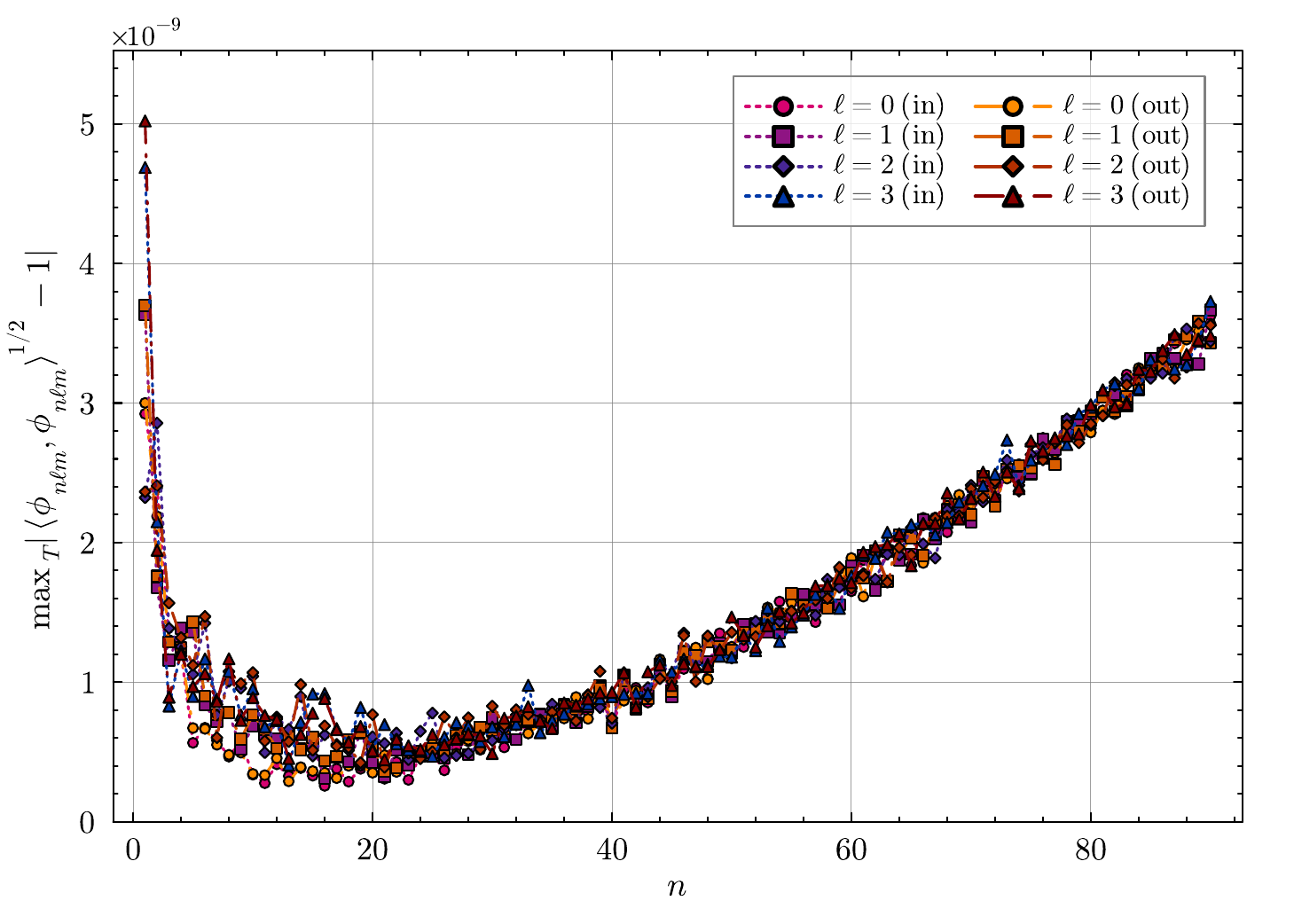}
  \end{subfigure}
  \caption{Left: Absolute error of the norm of representative in (solid) and out (dashed) modes, computed at different times for selected $(n,\ell)$.
Right: Maximum absolute deviation over time of the mode norms as a function of $n$. The inner products are evaluated using Eq.~\eqref{eq:inner-prod-oscillating-coords}. Numerical errors are bounded by $5\times 10^{-9}$, indicating a high degree of numerical stability and accuracy.}
  \label{fig:time-evolution-modes-Schwarzschild-norm}
\end{figure}

The results for the squared norm of several representative modes are shown in Fig.~\ref{fig:time-evolution-modes-Schwarzschild-norm} as a function of time $T$. The maximum error for each mode, used to obtain subsequent results, is also depicted in Fig.~\ref{fig:time-evolution-modes-Schwarzschild-norm}. Although the normalization is not strictly constant due to numerical errors, the deviations are bounded by $5 \times 10^{-9}$, indicating a high degree of numerical stability and accuracy. As expected, modes with larger values of $n$ exhibit slightly larger deviations, since their rapid oscillatory behavior, approximately $e^{-i\omega_{n\ell}T}$, makes them more sensitive to time-integration errors.

We now have all the ingredients to compute the Bogoliubov coefficients and, therefore, quantify particle creation induced by the stellar oscillations.

\subsection{Spectrum $| \beta_{\vec{k}\vec{k}'}|$ and total particle number $\langle \rm in | N_{\rm out} | \rm in \rangle$ \label{sec:calculation-beta}}

\begin{figure}[!htbp]
  \centering
  \includegraphics[width=0.6\linewidth]{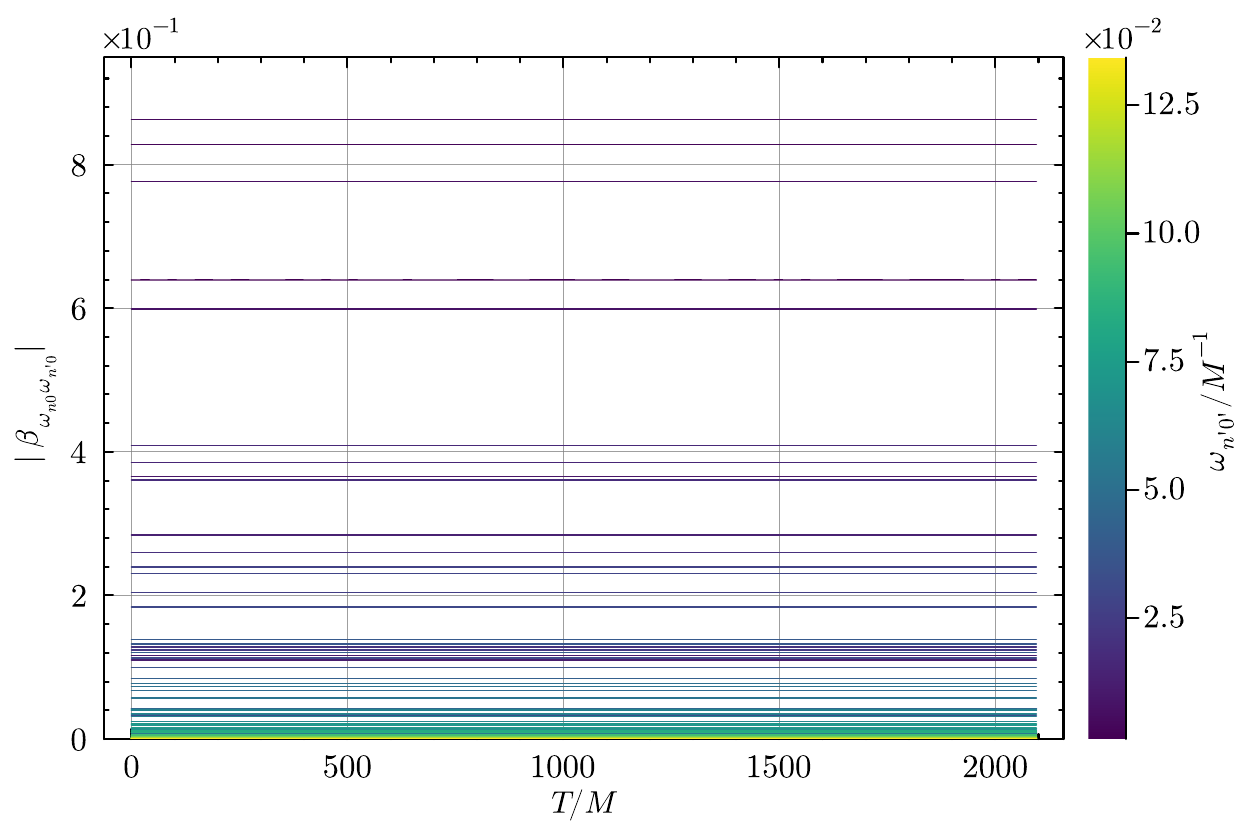}
      \caption{Numerical values of the Bogoliubov coefficients $\beta_{n\ell,n'\ell}$ for $\ell = 0$, $n = 20$ (corresponding to $\omega_{20, 0} = 2.98 \times 10^{-2} \approx \Omega$), and $n' = 1,\dots,90$, shown as a function of time $T$. The time-independence is verified with high accuracy. The colorbar indicates the frequency $\omega'$ of the out modes. }
  \label{fig:beta-l0-Schwarzschild-ground-state}
\end{figure}

In  Fig.~\ref{fig:beta-l0-Schwarzschild-ground-state}  we present the numerical results for $\beta_{n\ell,n'\ell}$     as a function of time $T$ for $\ell=0$, $n=1$ and $n'$ running up to 39, computed using (\ref{betacoefficients}). The results found  remain constant in time $T$ within numerical accuracy, as expected. 

\begin{figure}[!htbp] 
  \centering
  \begin{subfigure}[b]{0.49\textwidth}
     \centering     \includegraphics[width=\columnwidth]{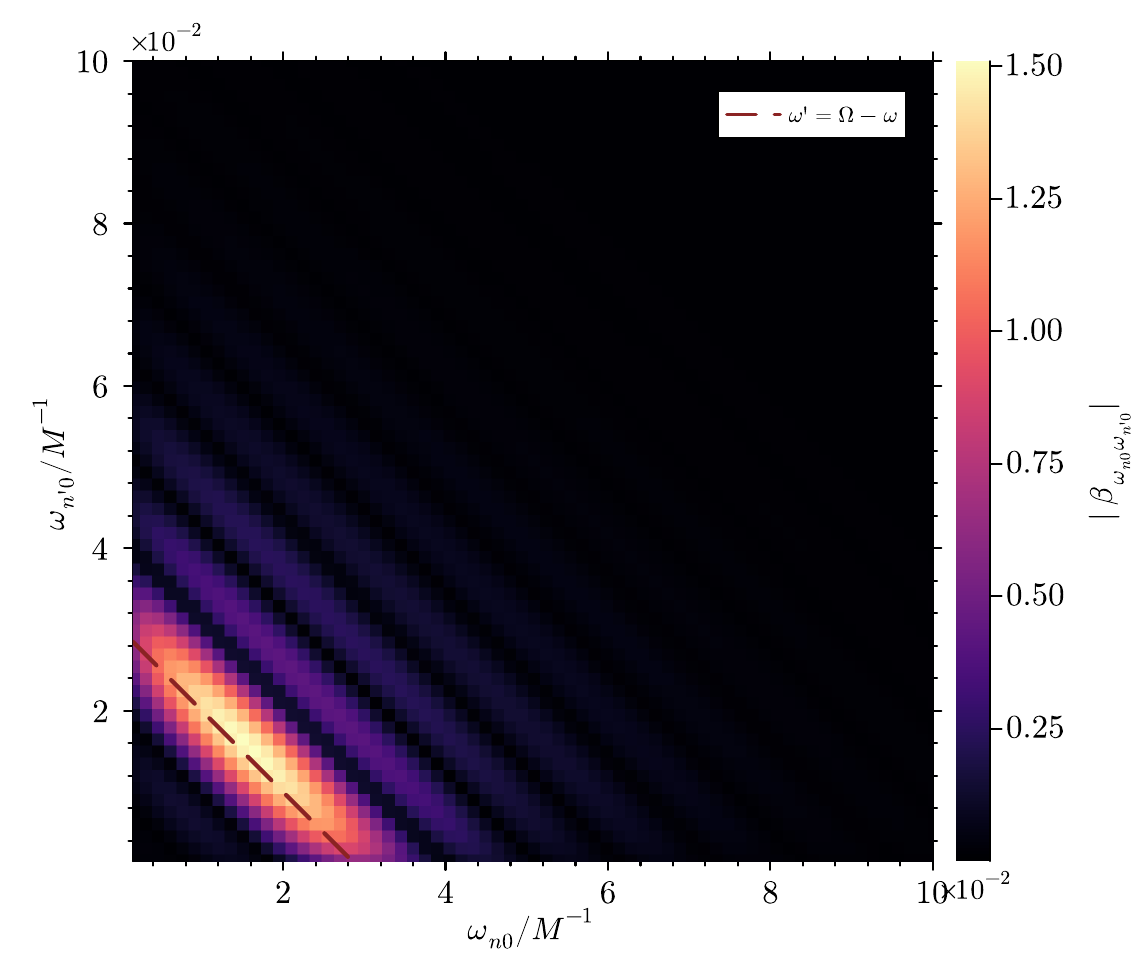}
  \end{subfigure}
  \hfill
  \begin{subfigure}[b]{0.49\textwidth}
     \centering
     \includegraphics[width=\columnwidth]{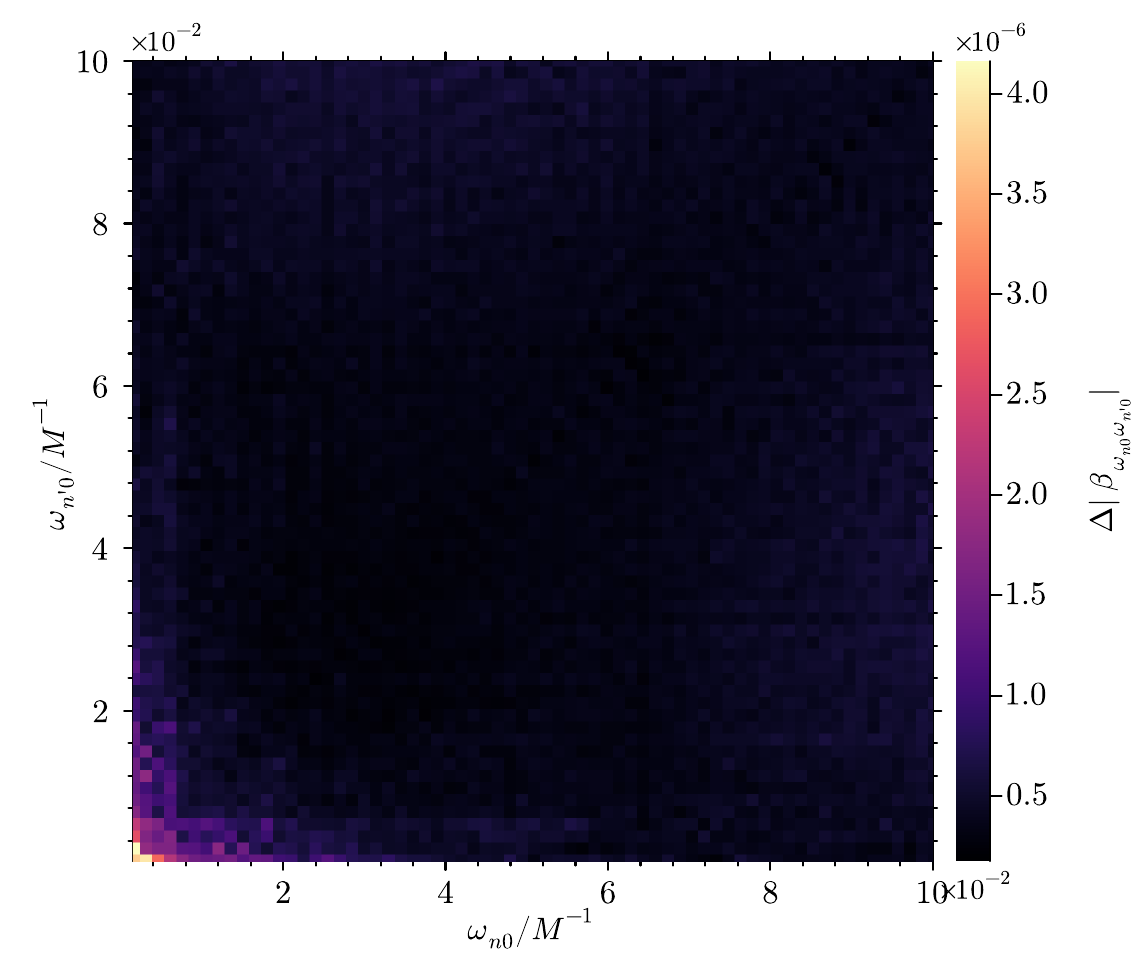}
  \end{subfigure}
  \begin{subfigure}[b]{0.49\textwidth}
     \centering     \includegraphics[width=\columnwidth]{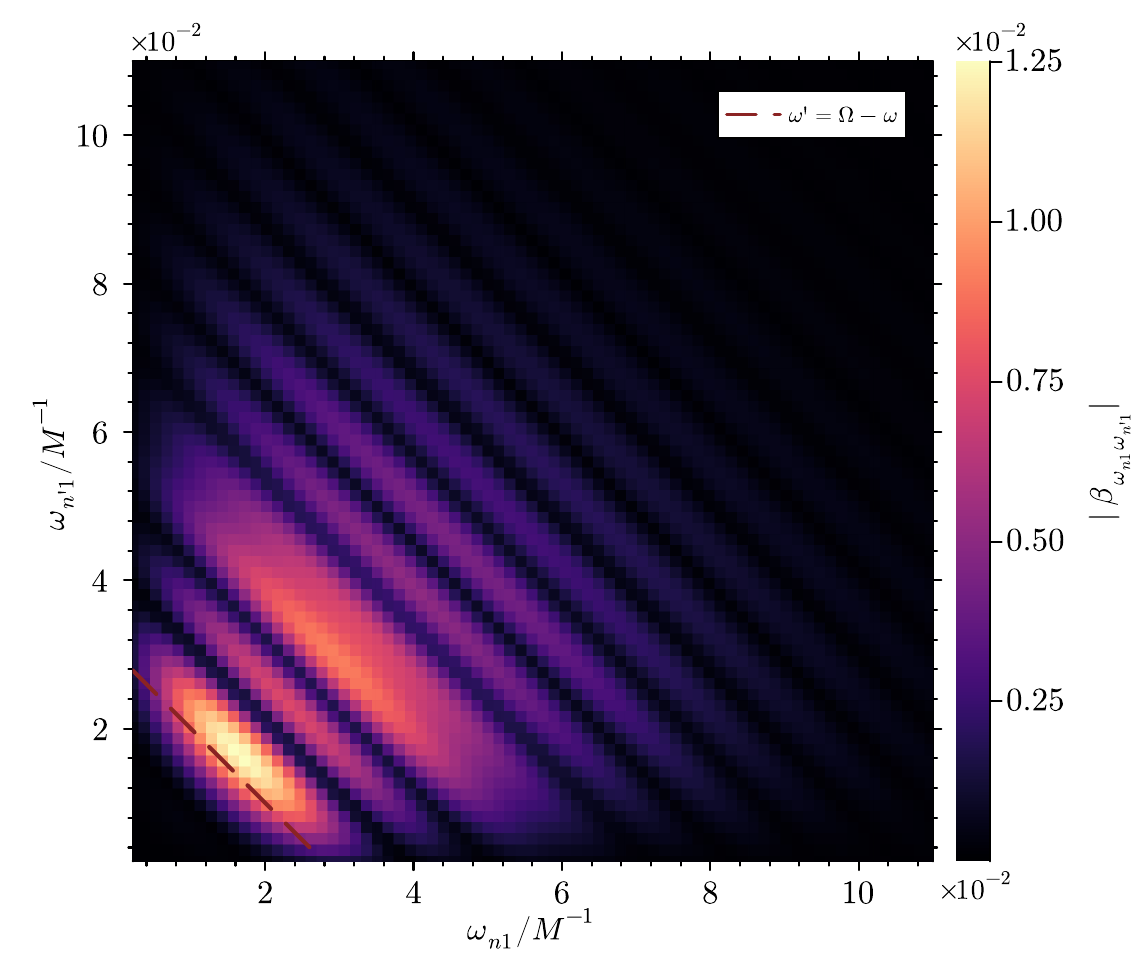}
  \end{subfigure}
  \hfill
  \begin{subfigure}[b]{0.49\textwidth}
     \centering
     \includegraphics[width=\columnwidth]{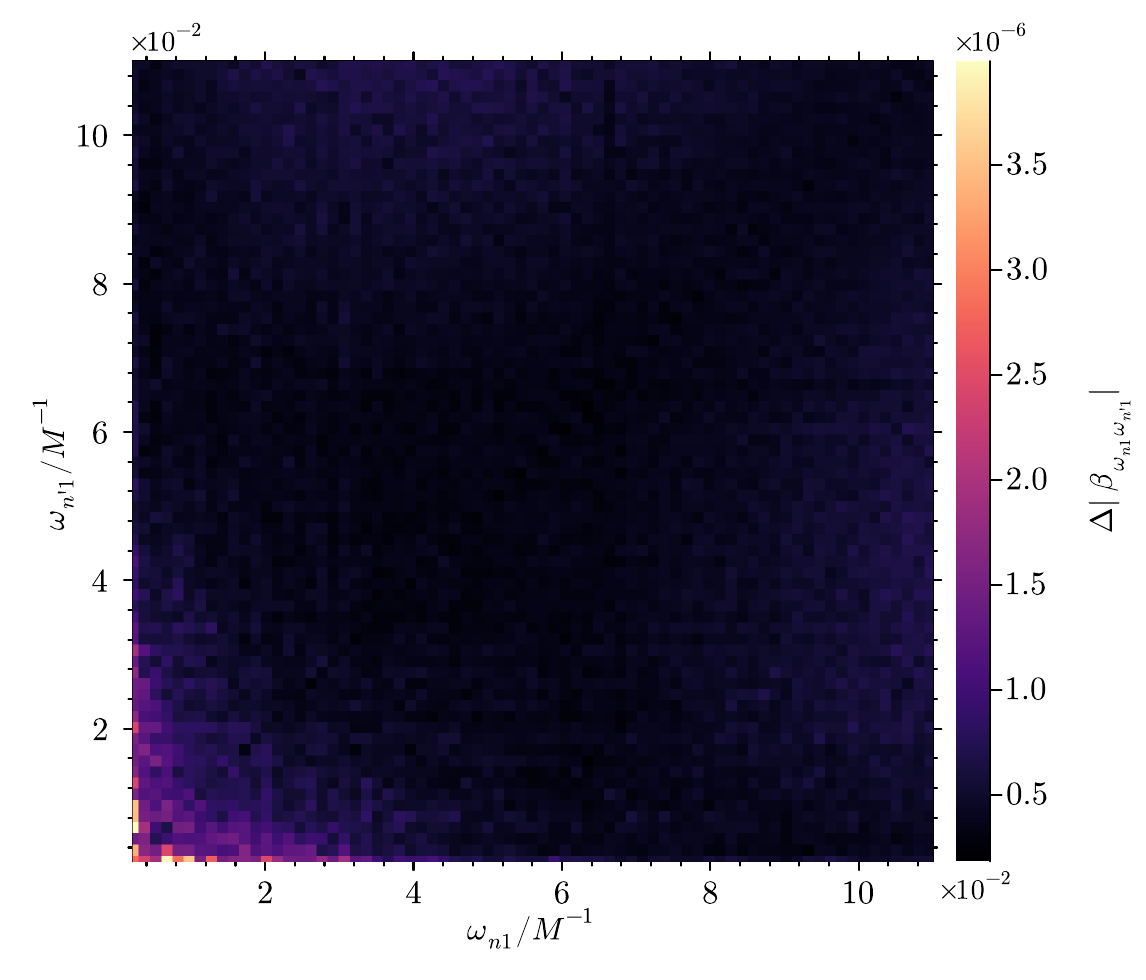}
  \end{subfigure}
  \begin{subfigure}[b]{0.49\textwidth}
     \centering     \includegraphics[width=\columnwidth]{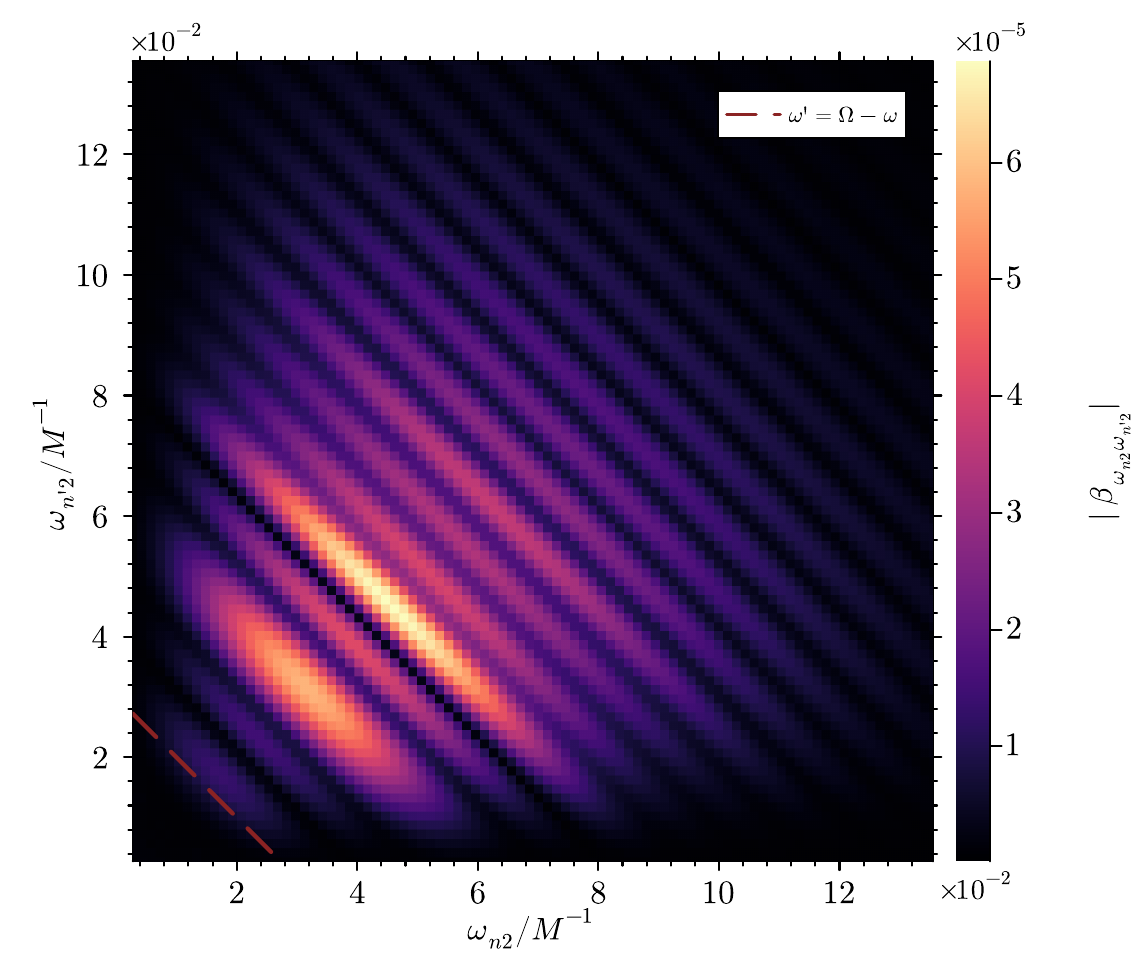}
  \end{subfigure}
  \hfill
  \begin{subfigure}[b]{0.49\textwidth}
     \centering
     \includegraphics[width=\columnwidth]{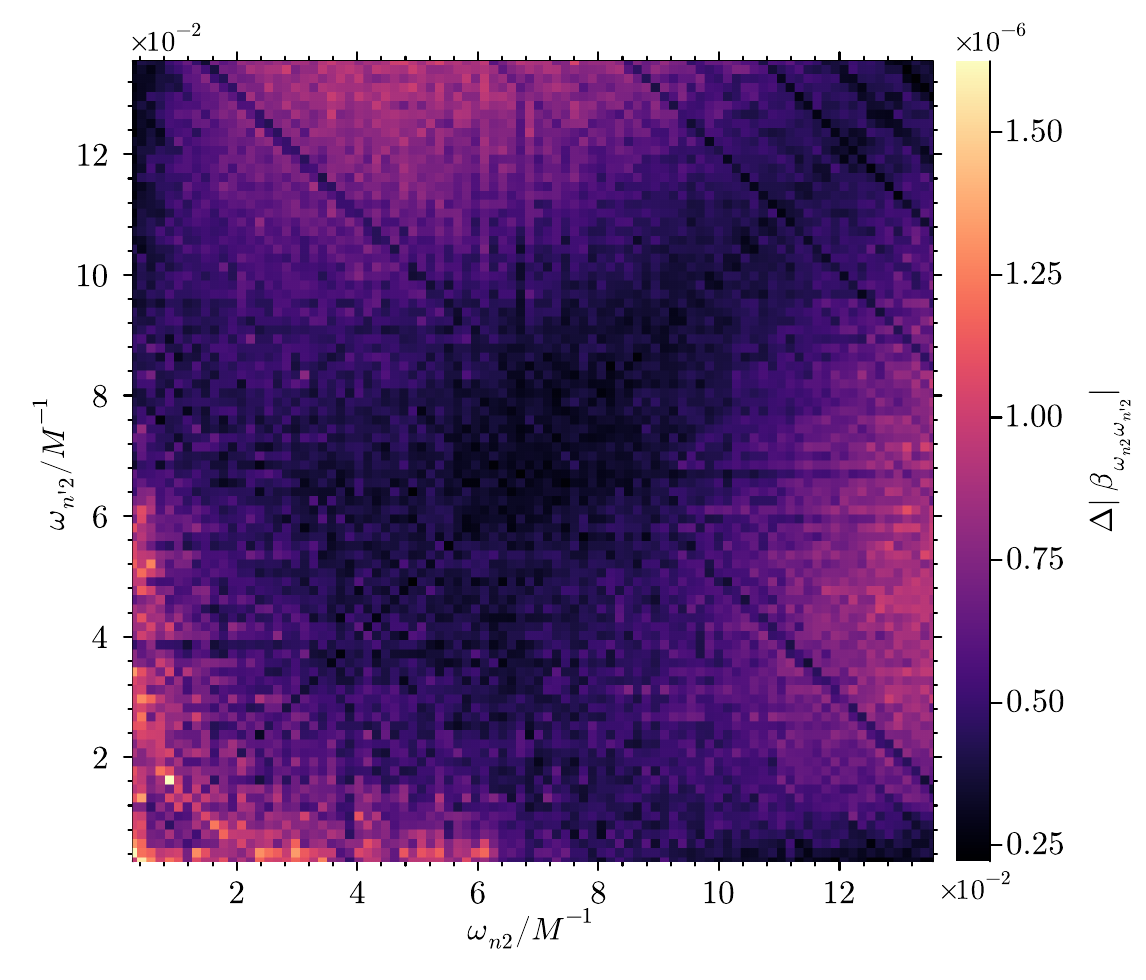}
  \end{subfigure}
  \caption{Particle spectra $|\beta_{n\ell,n'\ell}|$ computed from Eq.~(\ref{betacoefficients}) for $\ell=0,1,2$ as functions of the ``in'' ($n$) and ``out'' ($n'$) mode indices. The left column displays the absolute value of the Bogoliubov coefficients, while the right column shows the corresponding numerical uncertainty, estimated using Eq.~(\ref{errorestimate}). From top to bottom, the rows correspond to $\ell=0$, $\ell=1$, and $\ell=2$, respectively. The dashed red line indicates the resonant condition $\Omega \simeq \omega_{n\ell} + \omega_{n'\ell}$.}
  \label{fig:resonance-2}
\end{figure}

In Fig.~\ref{fig:resonance-2} we display the numerical values obtained for $|\beta_{n \ell, n' \ell}|$ as a function of both the in- and out-mode frequencies as well as their numerical error for $\ell =0,1,2$. The $|\beta_{n \ell, n' \ell}|$ plot is constructed by averaging the value obtained for $|\beta_{n \ell,n' \ell}|$ over all times $T$. The numerical error  is estimated as 
\bea
\Delta |\beta_{n \ell, n' \ell}|:=\frac{\text{max}_{T\in[T_0,T_{\rm max}]}|\beta_{n \ell, n' \ell}| - \text{min}_{T\in[T_0,T_{\rm max}]}|\beta_{n \ell, n' \ell}|}{2}\, ,\label{errorestimate}
\eea
i.e. half the difference between the maximum and minimum recorded values of $|\beta_{n \ell, n' \ell}|$ across all time $T$.

For $\ell=0$, the plot is invariant under the exchange  $(n,n')\to (n',n)$, and shows that $|\beta_{n0,n'0}|$ decreases for increasing $n$ or $n'$. More importantly, it exhibits a clear resonance  aligned along the diagonal $\Omega = \omega_{n0} + \omega_{n'0}$, for any $n,n'\in \mathbb N$, indicating that particle creation is strongly enhanced when this condition is satisfied. This is a physically expected behaviour: indeed, if we regard $\hbar \omega_{n\ell}$  and  $\hbar \omega_{n'\ell}$ as the energy of the two quanta of the particle pair excited by the oscillating spacetime, one expects particle creation is more efficient whenever the oscillating star provides energy $\hbar\Omega$ enough for exciting these two quanta.

The results shown in  Figs.~\ref{fig:resonance-2}, up to $\ell=2$, make manifest that the overall amplitude of the spectra  decreases considerably as we increase the values of $\ell$. Again, for each $\ell$ the the spectrum $|\beta_{n\ell,n'\ell}|$ has the mirror symmetry $|\beta_{n\ell,n'\ell}|=|\beta_{n'\ell,n\ell}|$, and the spectra tend to zero for increasing $n$ or $n'$, particularly when $\omega_{n\ell}, \omega_{n'\ell}>>\Omega$. Since the spectrum $\omega_\ell(n)\equiv \omega_{n\ell}$ of allowed frequencies  get shifted towards higher values as $\ell$ increasses, as shown in Fig.~\ref{fig:allowed-momenta-spectrum}, the resonance condition $\omega_{n'\ell} \simeq \Omega - \omega_{n\ell}$, depicted in Fig.~\ref{fig:resonance-2} for $\ell=0$, is shifted toward smaller values of $n'$ as $\ell$ increases. 

From the error estimates shown in Fig.~\ref{fig:resonance-2} we infer a numerical precision of order $10^{-6}$, which is sufficient to faithfully capture the spectrum $|\beta_{n \ell, n' \ell}|$  for $\ell \leq 2$. This limitation does not affect the computation of the total particle number, as we discuss below.

\begin{figure}[!htbp]
  \centering
  \begin{subfigure}[b]{0.49\textwidth}
     \centering \includegraphics[width=\columnwidth]{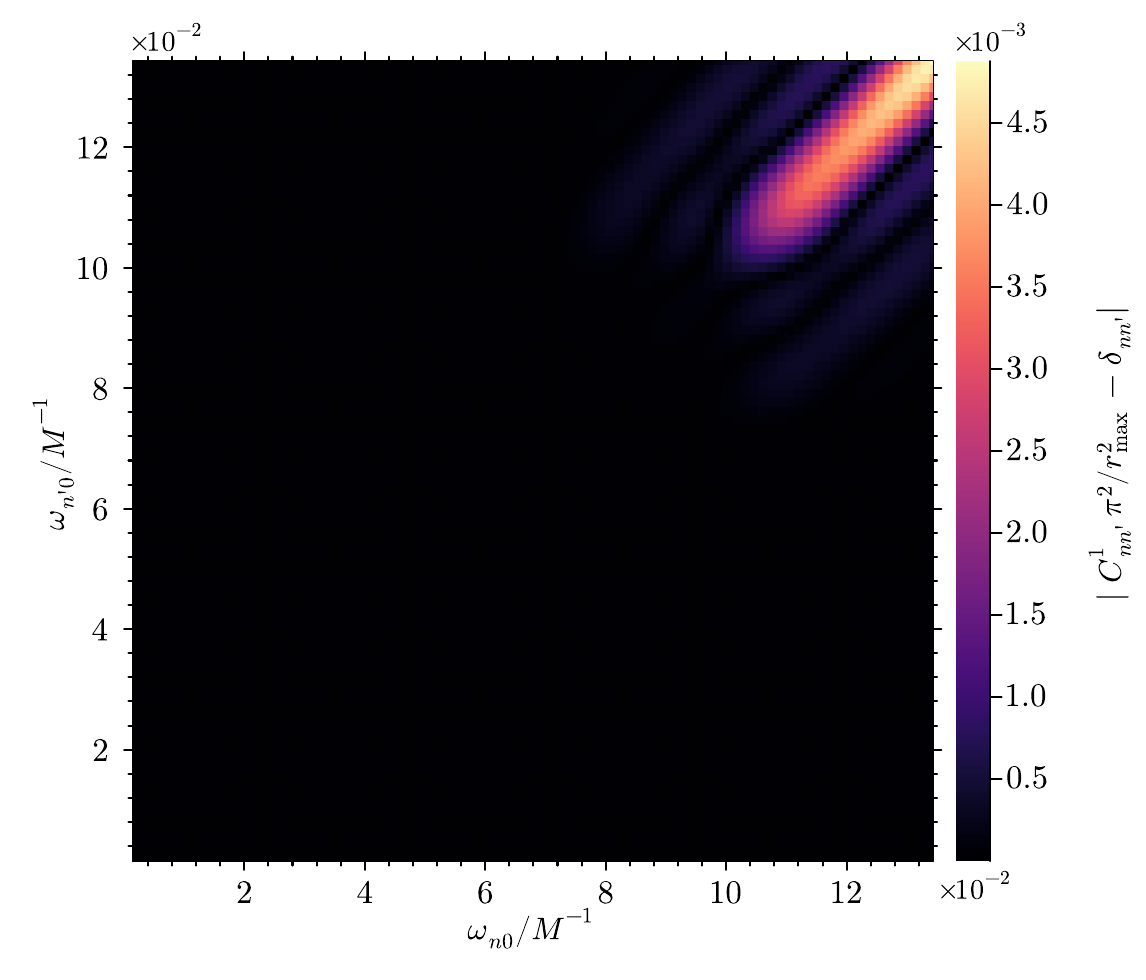}
  \end{subfigure}
  \hfill
  \begin{subfigure}[b]{0.49\textwidth}
     \centering
     \includegraphics[width=\columnwidth]{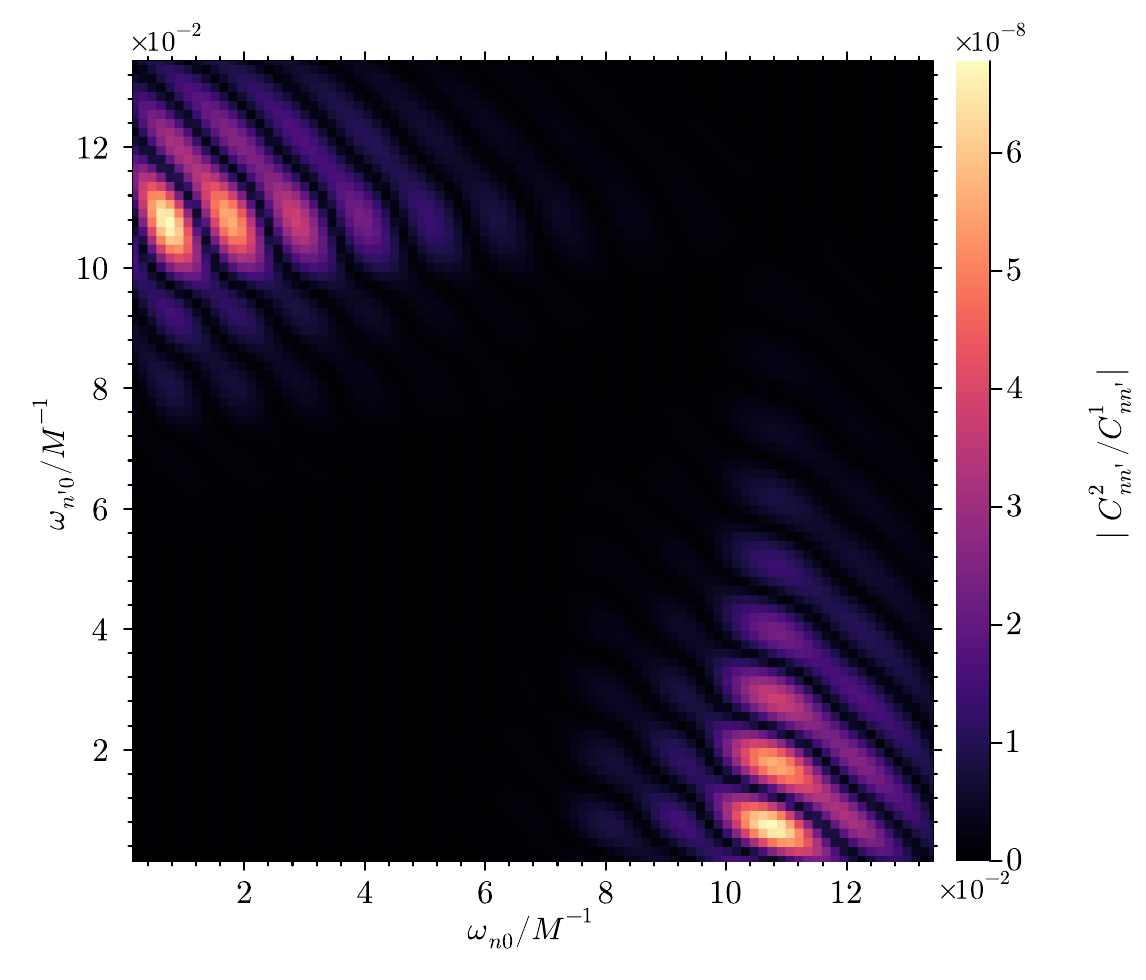}
  \end{subfigure}
  \begin{subfigure}[b]{0.49\textwidth}
     \centering \includegraphics[width=\columnwidth]{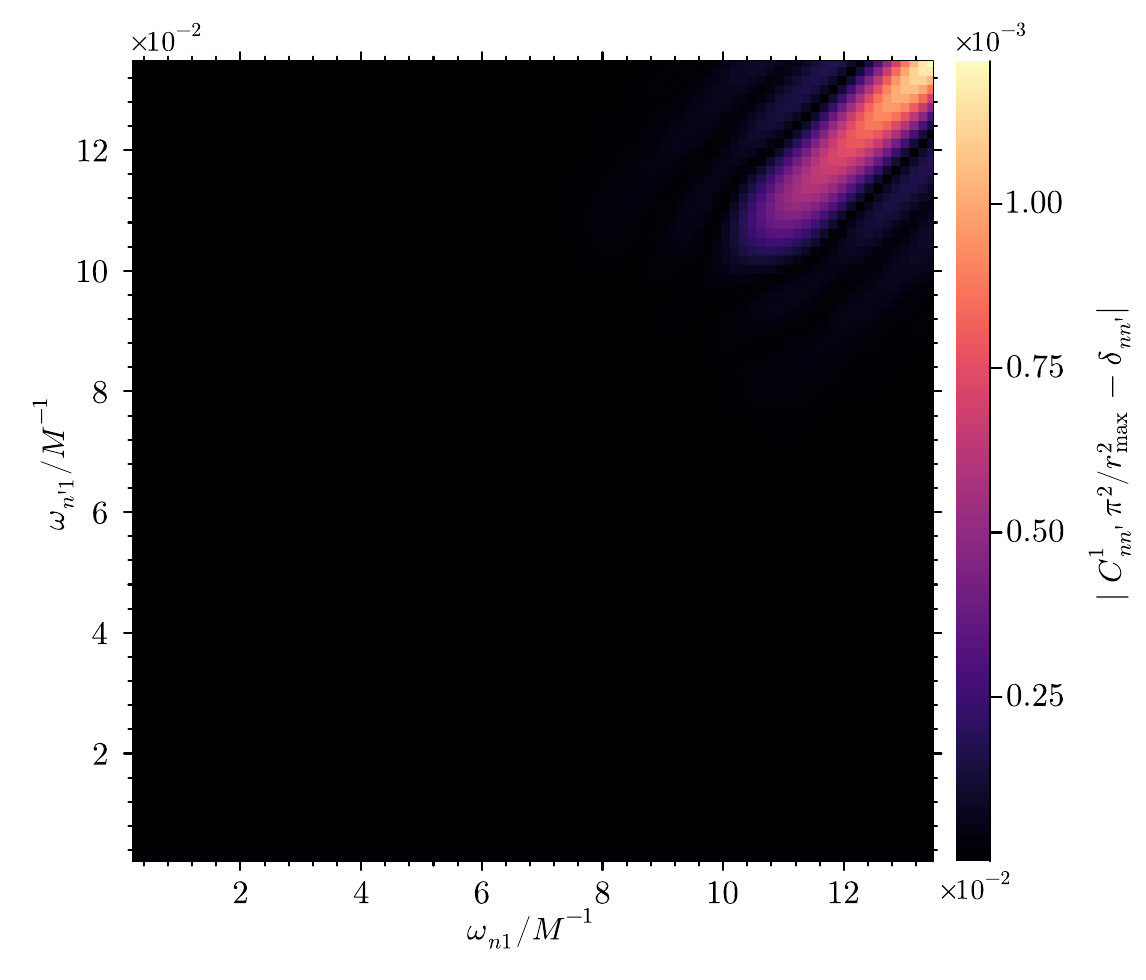}
  \end{subfigure}
  \hfill
  \begin{subfigure}[b]{0.49\textwidth}
     \centering
     \includegraphics[width=\columnwidth]{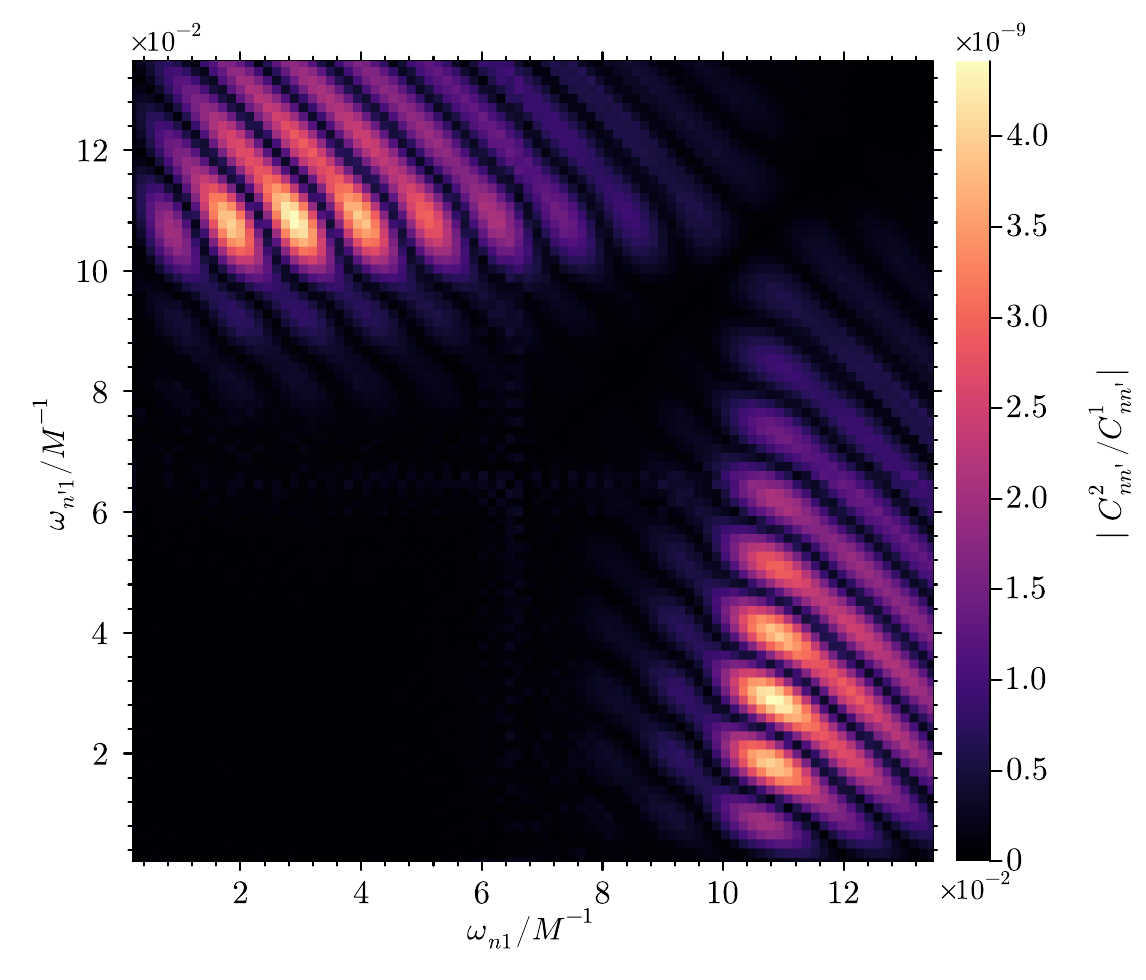}
  \end{subfigure}
  \begin{subfigure}[b]{0.49\textwidth}
     \centering \includegraphics[width=\columnwidth]{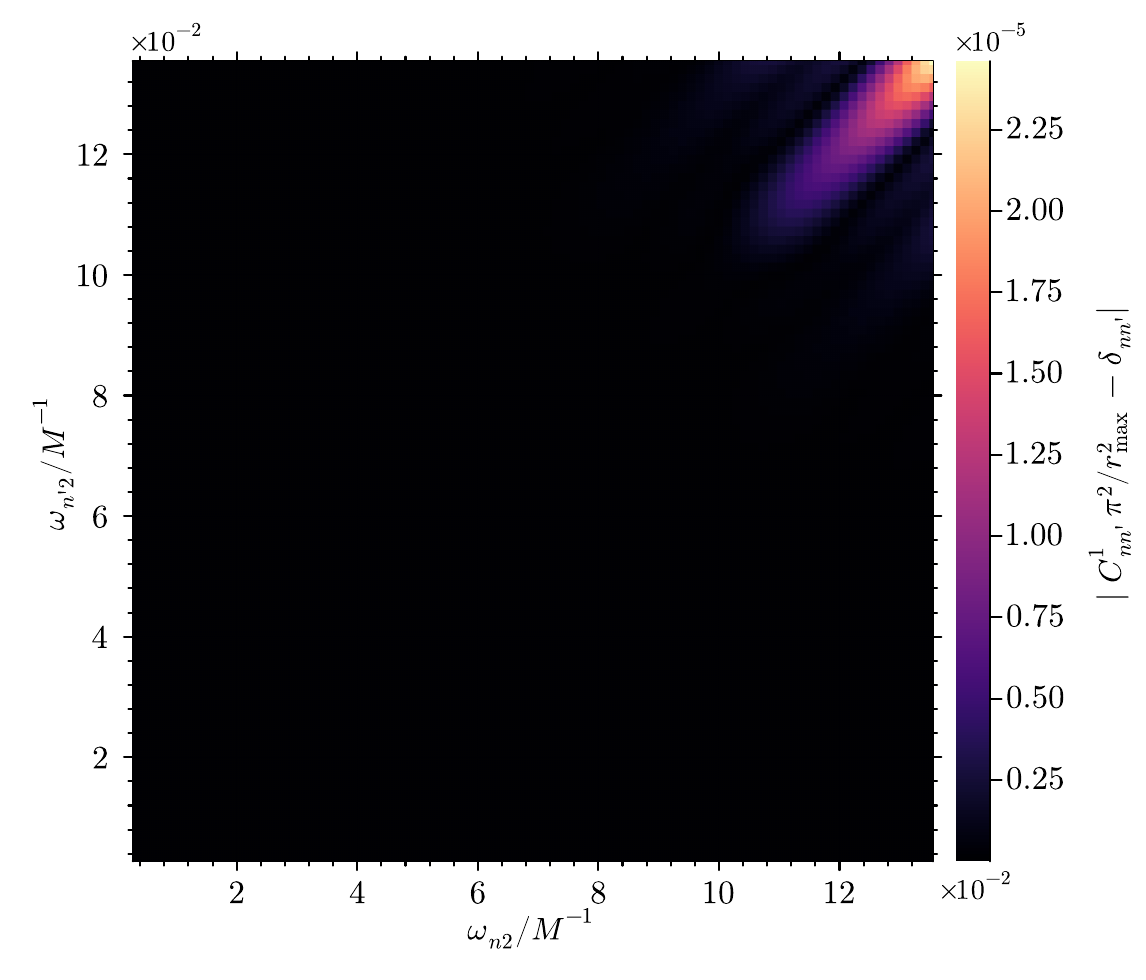}
  \end{subfigure}
  \hfill
  \begin{subfigure}[b]{0.49\textwidth}
     \centering
     \includegraphics[width=\columnwidth]{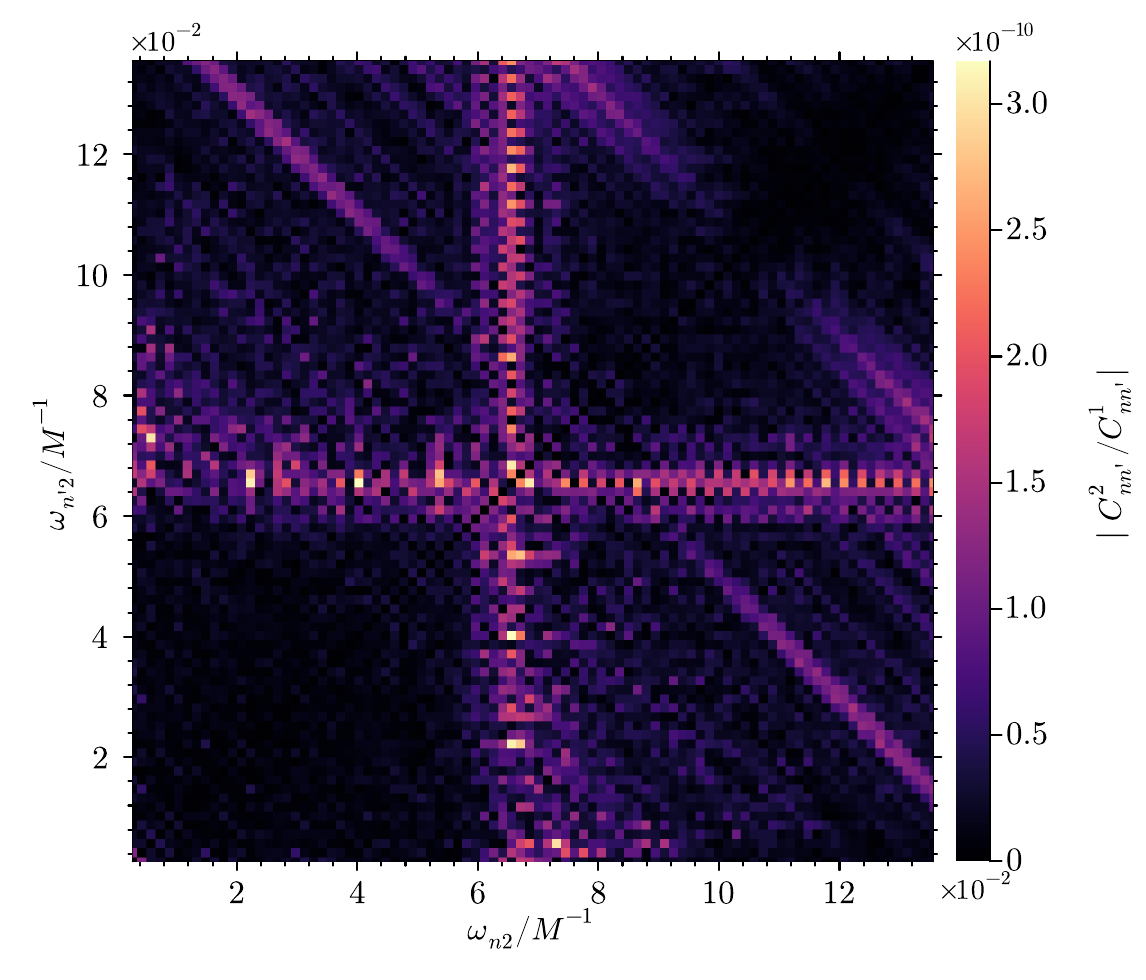}
  \end{subfigure}
  \caption{Unitarity conditions $C^{1,\ell}_{nn'}$ and $C^{2,\ell}_{nn'}$, computed from Eqs.~\eqref{constraint1} and \eqref{constraint2} for $\ell=0,1,2$, shown as functions of the ``in'' ($n$) and ``out'' ($n'$) mode indices.  The three panels on the left display the residuals $\frac{\pi^2}{r_{\rm max}^2}C^{1,\ell}_{nn'} - \delta_{nn'}$, while the three on the right show the relative (dimensionless) error $C^{2,\ell}_{nn'}/C^{1,\ell}_{nn}$, all of which are expected to vanish identically.  Rows one through three correspond to $\ell=0$, $\ell=1$, and $\ell=2$, respectively. }
  \label{fig:unitarity}
\end{figure}

To assess the reliability of our numerical results, we test the unitarity constraints (\ref{constraint1})-(\ref{constraint2}) for each value of $\ell$. Fig.~\ref{fig:unitarity} shows the numerical results for $C_{nn'}^{1,\ell}$ and $C_{nn'}^{2,\ell}$ defined in \eqref{constraint1} and \eqref{constraint2} respectively, for $\ell=0,1,2$. As one can appreciate, the $C_{nn'}^{1,\ell}$ condition is fulfilled to the $10^{-3}$ level, with largest  deviations from unitarity occurring at high values of $n$ and $n'$, where numerical errors are expected to accumulate. The $C^{2,\ell}_{nn'}$ condition is normalized with respect to $C_{nn}^{1,\ell}$ in order to remove the dimensions of the $\alpha$ and $\beta$ coefficients. The ratio $C^{2,\ell}_{nn'}/C^{1,\ell}_{nn}$  is zero up to numerical deviations of $10^{-8}$. Overall, these plots provide a strong consistency check of our numerical implementation.  

Interestingly, the qualitative functional form of $\beta_{n\ell,n'\ell}$ obtained numerically in Figs.~\ref{fig:resonance-2}  can be guessed on  physical grounds. Since it is known that a non-zero value of $\beta_{n\ell,n'\ell}$  arises solely from the acceleration dynamics of the background, and the only time-dependent contribution in our setup comes from the stellar surface  (\ref{stellaroscillation}), it is natural to expect a close connection between $\beta_{n\ell,n'\ell}$ and the Fourier transform $\cmcal F$ of the surface acceleration $\frac{d^2R_0(t)}{dt^2}$. This idea  is also supported by the decaying band structure in Figs.~\ref{fig:resonance-2}. Furthermore, motivated by the mirror symmetry observed in Figs.~\ref{fig:resonance-2} about the line $n'=n$, one is led to evaluate the Fourier transform at the frequency  $\omega_{n\ell} + \omega_{n'\ell}$, yielding    
\bea
|\beta_{n\ell,n'\ell}| \propto \left| \cmcal{F}\left\{\frac{d^2\,M \varepsilon(t) \sin(\Omega t)}{dt^2} \right\}(\omega_{n\ell} + \omega_{n'\ell}) \right|  = \frac{M}{2}(\omega_{n\ell} + \omega_{n'\ell})^2\left| \tilde{\varepsilon}( \omega_{n\ell} + \omega_{n'\ell} - \Omega) - \tilde{\varepsilon}( \omega_{n\ell} + \omega_{n'\ell} + \Omega) \right|\, ,
\eea
where $\cmcal{F}(\varepsilon)(\omega) = \tilde{\varepsilon}(\omega)$ denotes the Fourier transform of the switching function $\varepsilon(t)$.   For the choice in (\ref{eq:windowing-function}), this Fourier transform can be computed in full closed form, yielding (see Appendix~\ref{app:FT-epsilon} for details)
\bea
 \tilde{\varepsilon}(\omega) = \sqrt{\frac{\pi}{2}}\frac{A \Delta  e^{-i\omega (T_{\text{on}} + T_{\text{off}})/2}}{1 - e^{-2(T_{\text{off}} - T_{\text{on}})/\Delta}} \frac{\sin\left(\frac{\omega (T_{\text{off}} - T_{\text{on}})}{2}\right)}{\sinh\left(\frac{\pi \omega \Delta}{2}\right)},\label{fourier}
\eea
thus, one could expect
\bea
|\beta_{n\ell,n'\ell}| \propto  \sqrt{\frac{\pi}{8}}\frac{  A  M \Delta (\omega_{n\ell} + \omega_{n'\ell})^2}{1 - e^{-2(T_{\text{off}} - T_{\text{on}})/\Delta}}      \left | \frac{e^{-i(\omega_{n\ell}+\omega_{n'\ell}-\Omega) T_{\rm on} }-e^{-i(\omega_{n\ell}+\omega_{n'\ell}-\Omega) T_{\rm off} }}{2\sinh\left(\frac{\pi (\omega_{n\ell}+\omega_{n'\ell}-\Omega) \Delta}{2}\right)}-  \frac{e^{-i(\omega_{n\ell}+\omega_{n'\ell}+\Omega) T_{\rm on} }-e^{-i(\omega_{n\ell}+\omega_{n'\ell}+\Omega) T_{\rm off} }}{2\sinh\left(\frac{\pi (\omega_{n\ell}+\omega_{n'\ell}+\Omega) \Delta}{2}\right)} \right| \label{exactfit}
\eea
The numerical results obtained for $\ell=0$ in Fig.~\ref{fig:resonance-2} are accurately described by this expression, with maximum relative deviations of order $10^{-2}$, provided we  correct the global prefactor $(\omega_{n\ell} + \omega_{n'\ell})^2\to \sqrt{\omega_{n\ell}\omega_{n'\ell}}$. 
For higher values of $\ell$, however, this ansatz no longer reproduces the numerical results with comparable accuracy. We attribute this discrepancy to the presence of $\ell$-dependent graybody factors in the overall proportionality coefficient, arising from the backscattering of the field modes by the curved background geometry.

To quantify the total particle production, we compute the expectation value $\langle {\rm in}|N_{\rm out}|{\rm in}\rangle$ according to Eq.~\eqref{particlecreation}, which reduces to the formula
\bea
\langle {\rm in}|N_{\rm out}|{\rm in}\rangle=\frac{\pi^2}{r_{\rm max}^2}\sum_{\ell=0}^{\infty}\sum_{n=1}^{\infty}\sum_{n'=1}^{\infty}(2\ell+1) |\beta_{n\ell,n'\ell}|^2 \label{totalN2}\, .
\eea
Since the Bogoliubov coefficients $\beta_{n\ell,n'\ell}$  rapidly decay for $\omega_{n\ell}, \omega_{n'\ell} \gg \Omega$, and since $\omega_{n\ell}$  increases monotonically with both $n$ and $\ell$ (see Fig.~\ref{fig:allowed-momenta-spectrum}),  the infinite sums can be safely truncated to appropriate finite domains. In particular, the sums in $n$, $n'$ can be truncated to the domains displayed in the spectra Fig.~\ref{fig:resonance-2}, for $\ell=0$, $\ell=1$ and $\ell=2$. This truncation gives
\bea
\sum_{n=1}^{n_0}\sum_{n'=1}^{n'_0} |\beta_{n0,n'0}|^2&=&(224.344572 \pm 7 \times 10^{-6})M^2\, ,\\
\sum_{n=1}^{n_1}\sum_{n'=1}^{n'_1} |\beta_{n1,n'1}|^2&=&(0.03163135\pm 9 \times 10^{-8})M^2\, ,\\
\sum_{n=1}^{n_2}\sum_{n'=1}^{n'_2} |\beta_{n2,n'2}|^2&=&(1.873\pm 0.001) \times 10^{-6} M^2\, ,\\
\sum_{n=1}^{n_3}\sum_{n'=1}^{n'_3} |\beta_{n3,n'3}|^2&=&(10\pm 7) \times 10^{-10}M^2\, ,
\eea
where the quoted uncertainties reflect estimated numerical error bounds. We have further verified that these results remain stable under small variations of the truncation parameters $n_\ell$ and $n'_\ell$, indicating that truncation errors are negligible.

The mean value decreases by approximately four orders of magnitude at each step in $\ell$, ensuring that the overall sum~\eqref{totalN2} is already convergent at $\ell=2$. The dominance of the $\ell=0$ sector is consistent with the fact that the background perturbation is purely radial and therefore couples most efficiently to the spherically symmetric modes of the quantum field. The $\ell=3$ contribution is below the numerical uncertainty associated with the dominant $\ell=0$ sector, and has subsequently been discarded, since it is negligible within our level of precision. The $\ell=2$ contribution remains barely significant due to the multiplicity factor $(2\ell+1)$ in Eq.~\eqref{totalN2}.

\begin{figure}[!htbp]
  \centering
  \includegraphics[width=0.6\linewidth]{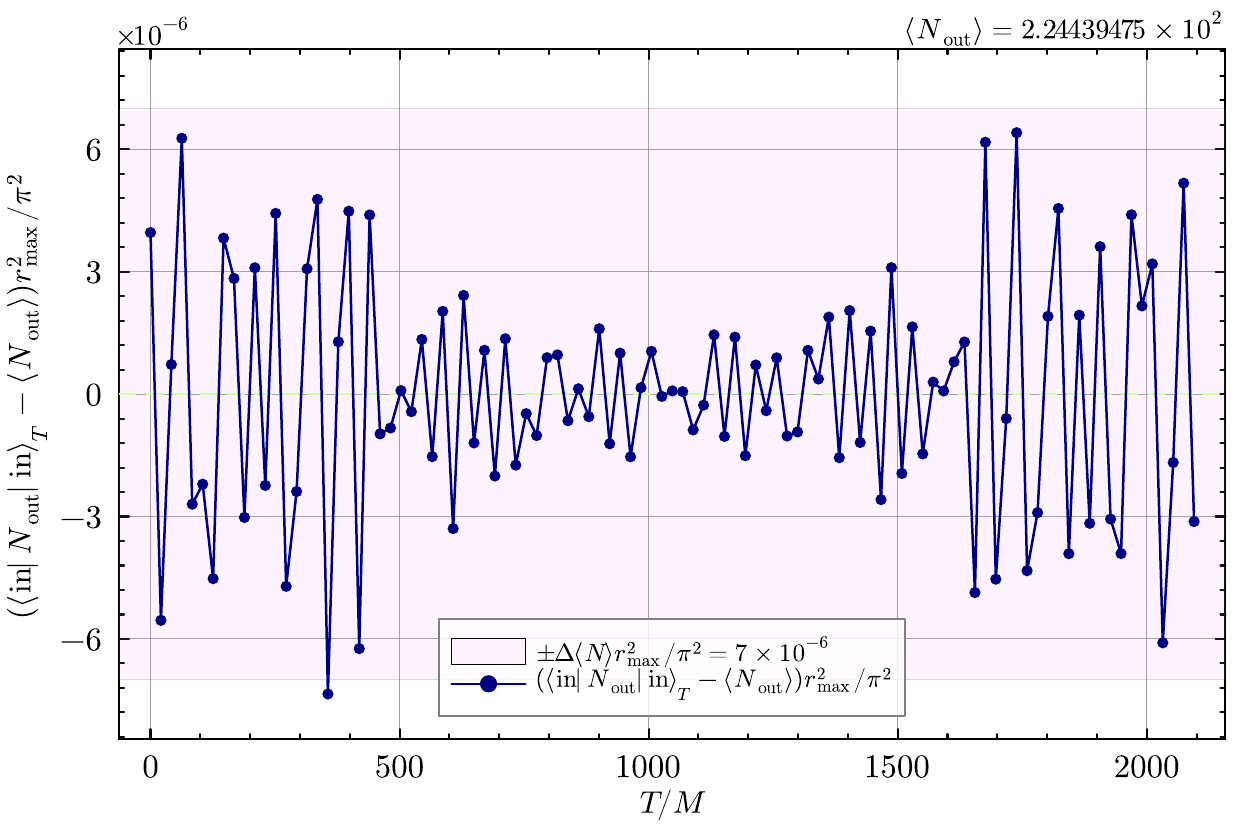}
  \caption{Numerical estimate of the total particle number $\langle {\rm in}|N_{\rm out}|{\rm in}\rangle$ as a function of time $T$, computed from the sum (\ref{totalN2}) truncated at $\ell=2$. The numerical value obtained  is  constant, exhibiting a maximum relative variation of $\lesssim 10^{-5}$.}
  \label{fig:sum-of-beta-norms}
\end{figure}

Using these  results, truncation of (\ref{totalN2}) at $\ell=2$ produces
\bea\label{totalN}
 \langle {\rm in}|N_{\rm out}|{\rm in}\rangle = \frac{\pi^2 M^2}{r_{\rm max}^2} (224.439475 \pm  7\times 10^{-6}) \sim 5\times 10^{-4}\, .
\eea
As shown in Fig \ref{fig:sum-of-beta-norms}, this result is  conserved in time $T$, exhibiting a maximum relative variation $\lesssim 10^{-5}\frac{\pi^2}{r_{\rm max}^2}$.

\section{Conclusions and future prospects}
\label{conclusions}

Particle pair creation from the quantum vacuum under extreme conditions is a well-established prediction of quantum field theory. This phenomenon appears in a variety of physical settings, ranging from electrodynamics (the Schwinger effect \cite{PhysRev.82.664, 1969JETP...30..660N, PhysRevD.2.1191, Dunne:2008kc}) to gravitational scenarios involving dynamical spacetimes. In this work, we have explored this quantum effect in a new and astrophysically relevant context, namely that of  oscillating compact stars.

Focusing on a free, massless, minimally coupled scalar field for simplicity, and adopting a toy model for the stellar oscillations, we have shown that the oscillatory motion of the stellar surface leads to spontaneous particle creation from the vacuum in the exterior spacetime. To this end, we have computed the Bogoliubov coefficients relating the in and out vacua using accurate numerical techniques, paying particular attention to issues of orthogonality and normalization of the field modes associated with the choice of boundary conditions. The computational framework is non-perturbative in nature, and thus applicable in the strong-field and fully relativistic regime. The resulting particle frequency spectrum, displayed  in Figs.~\ref{fig:resonance-2}  for different values of $\ell$, exhibits a distinctive resonant structure. Remarkably, this numerically obtained spectrum can be understood on simple physical grounds, allowing us to infer an analytical estimate for the case of $\ell=0$, given in Eq.~(\ref{exactfit}), which fits the numerical data reasonably well. The total number of particles created, summed over all frequencies and orbital angular momenta, has been computed in Eq.~(\ref{totalN}).

A central result of this work is the identification of a resonance condition when the sum of the frequencies of a particle pair matches the oscillation frequency of the background star, $\Omega \simeq \omega_{n0} + \omega_{n'0}$. This behavior admits a clear physical interpretation. The creation of a massless particle with frequency $\omega_{n\ell}$ requires the transfer of an energy quantum $\hbar \omega_{n\ell}$ from the background. Since particles are produced in pairs, efficient excitation occurs only when the stellar oscillations can supply the combined energy of the pair. Moreover, because the dynamical perturbation considered here is purely radial and preserves spherical symmetry, mode mixing occurs only within fixed $\ell$ sectors, and the s-wave sector ($\ell=0$) dominates the particle production, consistently with our numerical results.

Resonant behavior of this type has also been reported in studies of quantum fields confined within  cavities with moving boundaries in flat spacetimes (see e.g. \cite{physics2010007, Juarez_2022} and references therein), often within perturbative treatments and under several dynamical restrictions  (such as small amplitudes, slow boundary motion, short-time limit, parametric-resonant approximations, or effective $1+1$-dimensional reductions). The present analysis is fully non-perturbative, carried out in $3+1$ dimensions, and does not rely on weak-field, slow-motion, or small-amplitude expansions. The computations are performed directly in the strong-field regime and remain valid for arbitrarily long oscillation times within the numerical domain.

In the present work we have restricted attention to relatively low values of $\Omega$, in order to ensure numerical convergence of the mode sums in~\eqref{totalN2}  with a manageable number of modes, given the computational cost of the simulations. The value adopted here, $\Omega = 0.03/M$, is nevertheless slightly smaller than the typical frequency of the fundamental radial oscillation mode of a neutron star with comparable compactness. For instance, realistic neutron-star models with $C \simeq 0.3$ ($R \simeq 6.7M$) generally exhibit fundamental radial frequencies of order $\sim\,$10 kHz,  corresponding to $\Omega_{\mathrm{rad},0} M \sim 0.1$ for a canonical $1.4\,M_\odot$ star \cite{1983ApJS...53...93G}.   Still, the qualitative features identified in this work are not expected to differ. For such $\Omega$ the total particle number is expected to be higher, in fact.

Similarly, the parameter $\Delta$, which in our effective description sets the characteristic timescale for the decay of the stellar oscillations and can be interpreted as controlling the imaginary part of an effective quasi-normal mode frequency, is likely larger for realistic neutron stars with comparable compactness. In physically realistic stellar models, the damping of radial oscillations is expected to be dominated by viscous dissipation mechanisms, leading to characteristic damping timescales typically much longer than ${O}(10^2)M$~\cite{1990ApJ...363..603C}. In the present work we have instead adopted a smaller value of $\Delta$, primarily to keep the computational cost of the simulations manageable, since increasing $\Delta$ would require significantly longer time evolutions and substantially higher numerical resolution. Again the qualitative features of particle creation are not expected to change and, in view of (\ref{exactfit}), the total particle number is also expected to be higher.

The framework developed here can be extended in several directions. First, the background model considered is deliberately simplified, as a first step towards obtaining qualitative and physically transparent results. While the stellar oscillations were described using a toy model, the numerical methodology and conceptual framework can be straightforwardly generalized to include more physically realistic oscillations obtained from solutions of Einstein's equations. This would require solving for the fully dynamical spacetime metric, for instance using polytropic equations of state for the stellar matter within numerical relativity. Such an extension would also allow one to describe the mode dynamics inside the star by imposing regularity at the center, and avoiding the use of  boundary conditions at the stellar surface to effectively model the  coupling between the field modes and the stellar matter. We plan to address this more complete scenario in future work.

In addition, studying particle creation inside the star and its associated backreaction may be relevant for assessing potential quantum-induced instabilities, in line to those studied in the classical regime \cite{PhysRevLett.127.191101}. In particular, it would be interesting to investigate whether, for certain ranges of physical parameters such as the stellar compactness, the cumulative effect of particle creation could trigger or accelerate gravitational collapse. For sufficiently compact (and exotic) configurations, for instance if $R_0<<3M$ \cite{Cardoso_2019}, some field modes may become long-lived or effectively trapped between the stellar surface and the Regge-Wheeler potential barrier. In such a scenario, repeated reflections combined with stellar oscillations could lead to a progressive transfer of energy into these modes, potentially amplifying quantum effects over time. Such effects could therefore place indirect constraints on the observability and viability of stable exotic compact objects, as in \cite{PhysRevD.107.085023}. 

Additionally, the particle spectrum obtained here can also be expected to be further enhanced due its stimulated counterpart \cite{PhysRevD.13.3176}. If the quantum field is initially prepared in a non-vacuum state, it is well-known that particle production is generically enhanced by stimulated emission, with a contribution proportional to the occupation number of the initial state, as discussed for instance in cosmological settings~\cite{PhysRevD.83.063526, Agullo:2011xv}. While the present work focuses on spontaneous particle creation for a scalar field, such fields may be regarded as toy models capturing qualitative features of more astrophysically realistic quantum fields, like the electromagnetic field. Given that neutron stars host macroscopic electromagnetic fields and large photon occupation numbers, it is conceivable that stimulated effects could significantly amplify the spontaneous production mechanism identified here. Despite this, stimulated particle creation does not modify the characteristic frequencies of the process~\cite{PhysRevD.13.3176}. Accordingly, the resonant frequencies identified in this work are expected to remain unchanged, although their amplitude may be significantly amplified. Assessing whether such amplification could lead to observable consequences in realistic settings would require extending the present framework to non-vacuum initial states and to spin-1 fields. 

It is further instructive to discuss the characteristic frequencies associated with the particle production identified in this work. The resonance condition $\omega_{n\ell}+\omega_{n'\ell}\simeq\Omega$ implies that the modes most efficiently excited have frequencies of the same order as the stellar oscillation frequency. In the simplest symmetric case $\omega_{n\ell}\simeq\omega_{n'\ell}\simeq\Omega/2$, each created quantum carries a frequency set by half the oscillation frequency of the star. For realistic neutron-star models, the fundamental radial oscillation frequency typically lies in the kilohertz range, which would correspond to quanta with frequencies of order $\sim 1\,$kHz. If the present mechanism were extended to electromagnetic fields, such frequencies would fall in the extremely low- or very low-frequency radio bands, well below the MHz-GHz range commonly associated with observed pulsar radio emission \cite{alma9940729610001401}.

This observation suggests that, in its simplest form, the  particle creation mechanism studied here is unlikely to produce directly observable high-frequency electromagnetic radiation. Nevertheless, the analysis provides a useful reference scale for identifying the spectral window in which the effect operates. More importantly, extensions of the present framework (for instance, to include non-linear field interactions, or to highly rotating compact objects exhibiting  oscillatory dynamics) could potentially shift the relevant frequencies or enhance the amplitude of the effect. Exploring these possibilities, and assessing their possible observational implications, provides several promising avenues for future research.

{\bf \em Acknowledgments.} 
We thank Ivan Agullo, Vitor Cardoso, Alessandro Fabbri, Dimitrios Kranas, Jose Navarro-Salas, Javier Olmedo and Nicolas Sanchis-Gual for  useful discussions and feedback. 
ADR acknowledges support through {\it Atracci\'on de Talento Cesar Nombela} grant No 2023-T1/TEC-29023, funded by Comunidad de Madrid (Spain); as well as   financial support  via the Spanish Grant  PID2023-149560NB-C21, funded by MCIU/AEI/10.13039/501100011033/ FEDER, UE.  PLO is supported by the INFN Research Project QGSKY.

\appendix

\section{Time independence of the symplectic structure} \label{app:time-independence-symplectic}

For completeness, and given its relevance in the main text, we provide here an explicit proof of the time independence of the symplectic structure (\ref{symp})  under the boundary conditions adopted in this work.

Let $\phi_1$ and $\phi_2$ be two solutions of the Klein-Gordon equation on the spacetime $(\mathbb M,g_{ab})$. The symplectic structure on a spacelike hypersurface $\Sigma_t$ can be written as
\bea
\Omega(\phi_1,\phi_2)=\int_{\Sigma_t} d\Sigma_a \, \omega^a(\phi_1,\phi_2) \, ,
\eea
where $d\Sigma_a = t_a \, d\Sigma$, with $t_a$  the future-directed unit normal to $\Sigma_t$, and $\omega^a(\phi_1,\phi_2) = \phi_1 \nabla^a \phi_2 - \phi_2 \nabla^a \phi_1$ is called the symplectic current associated with the solutions $\phi_1$ and $\phi_2$. Using the Klein-Gordon equation, it follows immediately that this current is conserved, $\nabla_a \omega^a(\phi_1,\phi_2)=0$.

To establish the independence of $\Omega$ with respect to the choice of hypersurface, consider two spacelike hypersurfaces $\Sigma_{t_1}$ and $\Sigma_{t_2}$ bounding a spacetime region $\mathcal V$, together with the timelike boundary $\mathcal B$ defined by the stellar surface and spatial infinity. Applying Gauss' theorem to the conserved current, we obtain
\bea
0=\int_{\mathcal V} d^4x \, \sqrt{-g} \, \nabla_a \omega^a=\int_{\partial \mathcal V} dS_a \, \omega^a \, .
\eea
The boundary $\partial \mathcal V$ consists of $\Sigma_{t_2}$, $-\Sigma_{t_1}$ and $\mathcal B$, hence
\bea
\Omega_{t_2}(\phi_1,\phi_2)-\Omega_{t_1}(\phi_1,\phi_2)+\int_{\mathcal B} dS_a \, \omega^a=0 \, .
\eea
Therefore, the symplectic structure is time independent provided that the flux of the symplectic current through the timelike boundary $\mathcal B$ vanishes. We now analyze the two boundary components separately:

\medskip

\noindent
1) At the moving stellar surface we impose Dirichlet boundary conditions, $\phi(t,R_0(t),\theta,\varphi)=0$. Since both $\phi_1$ and $\phi_2$ vanish identically on the boundary, the symplectic current $\omega^a(\phi_1,\phi_2)$ vanishes there as well, and therefore no symplectic flux crosses the stellar surface.

\medskip

\noindent
2) At large $r$ we demand that the field modes behave asymptotically as their Minkowskian counterpart,
\bea
\phi(t,r,\theta,\varphi)
\sim
\sum_{\ell m}\int d\omega \,
A_{\omega \ell} j_\ell(\omega r)
e^{-i\omega t}
Y_{\ell m}(\theta,\varphi) \, .
\eea
for suitable $A_{\omega \ell}$.
Since $j_\ell(\omega r) \sim \sin(\omega r -\ell\pi/2)/r$ for large $r$, the radial component of the symplectic current decays as $1/r^2$ up to oscillatory factors. The surface element contributes a factor $r^2$, so the boundary flux reduces to oscillatory integrals in the frequency domain. These vanish in the limit $r\to\infty$ by virtue of the Riemann--Lebesgue lemma, assuming $A_{\omega\ell}$ decays faster than $1/\omega$ for large $\omega$.

\medskip

\noindent Since the symplectic flux vanishes both at the stellar surface and at spatial infinity, we conclude that
\bea
\Omega_{t_2}(\phi_1,\phi_2) = \Omega_{t_1}(\phi_1,\phi_2) \, ,
\eea
for arbitrary $t_1$ and $t_2$. The symplectic structure is therefore independent of the choice of spacelike hypersurface.

\section{Fourier transform of the switching function} \label{app:FT-epsilon}

In this Appendix we provide a proof of formula (\ref{fourier}). We start from the expression for the switching function as per Eq.~\eqref{eq:windowing-function}
\bea
\epsilon(T) = \frac{A}{4}\left[1 + \tanh\!\left(\frac{T - T_{\rm on}}{\Delta}\right)\right] \left[ 1 - \tanh\!\left(\frac{T - T_{\rm off}}{\Delta}\right) \right] \, ,
\eea
and introduce the dimensionless variables $a := \frac{T_{\rm on}}{\Delta}$, $b := \frac{T_{\rm off}}{\Delta}$,  $x := \frac{T}{\Delta}$, so that $T - T_{\rm on} = \Delta (x-a)$ and $T - T_{\rm off} = \Delta (x-b)$. Using the identities $1 \pm \tanh(u) = \frac{e^{\pm u}}{\cosh(u)}$, the function can be written compactly as
\bea
g(x) := \epsilon(\Delta x)=\frac{A}{4}\frac{e^{\,b-a}}{\cosh(x-a)\cosh(x-b)} \, .
\eea

Now, we adopt the unitary convention for the Fourier transform,
\bea
\cmcal{F}\{\epsilon(T)\}(\omega)=\tilde \epsilon(\omega)=\frac{1}{\sqrt{2\pi}} \int_{-\infty}^{\infty} dT\, \epsilon(T)\, e^{-i\omega T} \, ,
\eea
which implies 
\bea
\tilde \epsilon(\omega)=\frac{\Delta}{\sqrt{2\pi}} \int_{-\infty}^{\infty} dx\, g(x)\, e^{-i\omega\Delta x} \, .
\eea
Defining the dimensionless frequency $k := \Delta \omega$, we obtain
\bea
\tilde \epsilon(k)= \frac{A\Delta e^{\,b-a}}{4\sqrt{2\pi}}\int_{-\infty}^{\infty} dx\, \frac{e^{-ikx}}{\cosh(x-a)\cosh(x-b)} \, .
\eea
Introducing the variable $u := x-a$ and defining $d := b-a$, this becomes
\bea
\tilde \epsilon(k)&=&\frac{A\Delta e^{\,d-ika}}{4\sqrt{2\pi}}\int_{-\infty}^{\infty} du\,\frac{e^{-iku}}{\cosh(u)\cosh(u-d)} \nonumber\\
&=&\frac{A\Delta e^{\,d-ika}}{4\sqrt{2\pi}\sinh(d)}\int_{-\infty}^{\infty} du\,e^{-iku} \left[ \tanh(u) - \tanh(u-d) \right] \, ,
\eea
where we used the identity
\bea
\frac{1}{\cosh(u)\cosh(u-d)}=\frac{\tanh(u) - \tanh(u-d)}{\sinh(d)} \, .
\eea
Using the translation property of the Fourier transform,
$\cmcal{F}\{\epsilon(u-d)\}(\omega)=e^{-i\omega d}\cmcal{F}\{\epsilon(u)\}(\omega)$,
we further obtain
\bea
\tilde \epsilon(k)=\frac{A\Delta e^{\,d-ika}}{4\sinh(d)}\left[\cmcal{F}\{\tanh(u)\}(k)-e^{-ikd}\cmcal{F}\{\tanh(u)\}(k)\right].
\eea

The Fourier transform of $\tanh(u)$ is well-defined in the sense of distributions and reads
\bea
\cmcal{F}\{\tanh(u)\}(k)=-\,i\sqrt{\frac{\pi}{2}}\,\mathrm{csch}\!\left(\frac{\pi k}{2}\right).
\eea
Substituting this expression and using the identities
\bea
1 - e^{-ikd} = 2i e^{-ikd/2} \sin\!\left(\frac{kd}{2}\right), \qquad \frac{e^{d}}{\sinh(d)} = \frac{2}{1-e^{-2d}} \, ,
\eea
we arrive at
\bea
\tilde \epsilon(k)=A\Delta \frac{\sqrt{\pi}}{\sqrt{2}(1-e^{-2d})} \frac{\sin(kd/2)}{\sinh(\pi k/2)}\,e^{-ik\left(a+\frac{d}{2}\right)} \, .
\eea
Finally, restoring the original variables $d = \frac{T_{\rm off}-T_{\rm on}}{\Delta}$, $a + \frac{d}{2} = \frac{T_{\rm on}+T_{\rm off}}{2\Delta}$,  and $k = \Delta \omega$, the Fourier transform in physical frequency units is
\bea
\tilde \epsilon(\omega) =\sqrt{\frac{\pi}{2}}\frac{A\Delta}{1-e^{-2(T_{\rm off}-T_{\rm on})/\Delta}} \frac{\sin\!\left[\omega (T_{\rm off}-T_{\rm on})/2\right]}{\sinh\!\left(\pi \Delta \omega /2\right)}\,e^{-i\omega (T_{\rm on}+T_{\rm off})/2}\, .
\eea

\section{Normalization factor $A_{n\ell}$}

Although only the combination $A_{n\ell} R_{n\ell}(z)$ is physically meaningful, it can be useful to visualize the normalization factor $A_{n\ell}$ separately. 

This requires specifying a criterion to fix the overall scaling of the numerically obtained eigenfunctions $R_{n\ell}(z)$. For definiteness, we normalize the eigenfunctions so that they match the corresponding Minkowski radial modes in the asymptotic region of the numerical domain, namely $R_{n\ell}(z) \propto j_{\ell}(\omega_{n\ell} z)$ for large $z$. In practice, the numerical solution is extended over an additional set of 500 grid points in the outer region, and a scaling factor $B_{n\ell}$ is determined by fitting  $B_{n\ell} R_{n\ell}(z) = j_{\ell}(\omega_{n\ell} z)$
within this asymptotic domain. The procedure is iterated until a relative fitting error below $10^{-5}$ is achieved. The eigenfunctions are then rescaled according to $R_{n\ell}(z) \rightarrow B_{n\ell} R_{n\ell}(z)$, ensuring consistency of both $A_{n\ell}$ and $R_{n\ell}$ with the $z \to \infty$ and $M \to 0$ limits.

\begin{figure}[!htbp]
  \centering  \includegraphics[width=0.65\columnwidth]{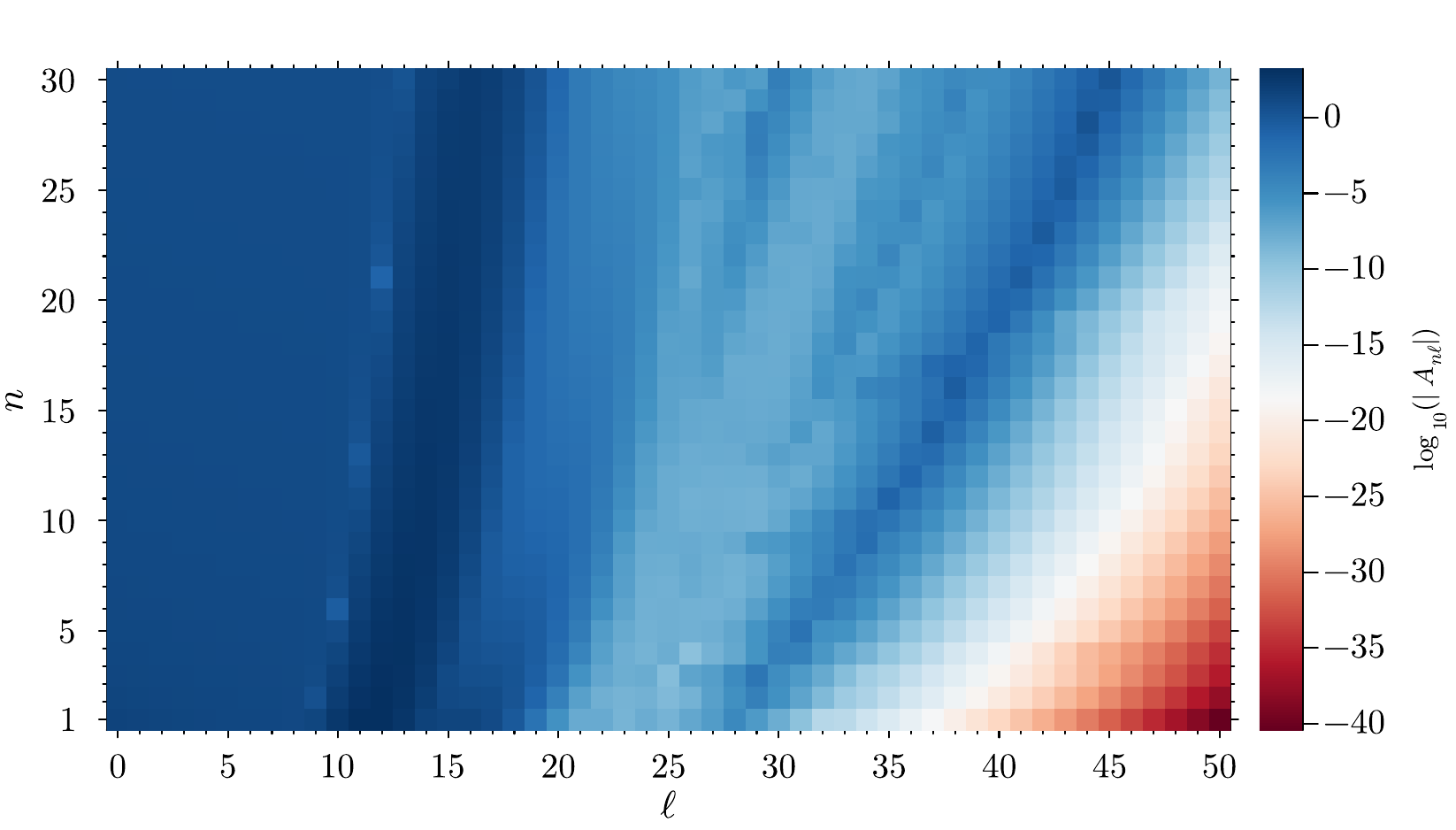}
  \caption{Values of $A_{n \ell}$ for $\ell \in [0, 50]$ and $n \in [1, 30]$ in logarithmic scale with respect to radial profiles satisfying $R_{ n \ell}(z) = j_{\ell}(\omega_{n \ell} z)$ as $z \to \infty$.} 
  \label{fig:A_nl_logscale}
\end{figure}

It is important to stress that the product $A_{n\ell} R_{n\ell}$ is independent of this choice of scaling, since the normalization factor $A_{n\ell}$ is computed using Eq.~\eqref{Akl-squared-early-times}. The resulting values of $A_{n\ell}$ are shown in Fig.~\ref{fig:A_nl_logscale}.

\bibliography{references}

\end{document}